\documentclass[prl,superscriptaddress,amsmath,amssymb,twocolumn,nofootinbib]{revtex4-2}
\usepackage{graphicx,bm}
\usepackage[dvipsnames]{xcolor}
\usepackage[colorlinks,bookmarks=false,citecolor=blue,linkcolor=Red,urlcolor=blue,
]{hyperref}
\usepackage[percent]{overpic}
\usepackage{bbm}
\usepackage{CJKutf8}
\usepackage[bbgreekl]{mathbbol}
\usepackage[normalem]{ulem}
\usepackage{braket}
\usepackage[ruled,vlined]{algorithm2e}
\usepackage{comment}
\usepackage{amsthm}
\usepackage{lipsum}
\usepackage{enumerate}
\usepackage{enumitem}
\usepackage{subfigure}
\graphicspath{{./figs/}}

\begin{document}

\title{Observable-Based Learning of Effective Entanglement Hamiltonians}

\author{Siyi Yang}
\thanks{These authors contributed equally to this work.}
\affiliation{State Key Laboratory of Surface Physics and Department of Physics, Fudan University, Shanghai 200438, China}
\affiliation{Department of Physics, School of Science and Research Center for Industries of the Future, Westlake University, Hangzhou 310030,  China}
\affiliation{Institute of Natural Sciences, Westlake Institute for Advanced Study, Hangzhou 310024, China}

\author{Yuzhi Liu}
\thanks{These authors contributed equally to this work}
\email{liuyuzhi17@mails.ucas.ac.cn}
\affiliation{AI for Science Institute, Beijing 100080, China}

\author{Yi-Ming Ding (\begin{CJK*}{UTF8}{gbsn}丁一茗\end{CJK*})}
\email{dingyiming@westlake.edu.cn}
\affiliation{State Key Laboratory of Surface Physics and Department of Physics, Fudan University, Shanghai 200438, China}
\affiliation{Department of Physics, School of Science and Research Center for Industries of the Future, Westlake University, Hangzhou 310030,  China}
\affiliation{Institute of Natural Sciences, Westlake Institute for Advanced Study, Hangzhou 310024, China}

\author{Zheng Yan}
\email{zhengyan@westlake.edu.cn}
\affiliation{Department of Physics, School of Science and Research Center for Industries of the Future, Westlake University, Hangzhou 310030,  China}
\affiliation{Institute of Natural Sciences, Westlake Institute for Advanced Study, Hangzhou 310024, China}


\begin{abstract}
Determining the entanglement Hamiltonian (EH) of an interacting quantum many-body system remains a fundamental challenge beyond special analytically tractable cases.
Here we develop an observable-matching framework for learning effective EHs, which we demonstrate using response estimates from quantum Monte Carlo.
As applications of this framework, we first calibrate the overall prefactor of the lattice Bisognano--Wichmann (LBW) construction using equal-time correlations and then learn spatially resolved couplings beyond its prescribed profile, without assuming a specific functional form for their spatial dependence.
Trained solely on local correlations, the learned EHs reproduce benchmarked von Neumann and second R\'enyi entropies in $1$D and second R\'enyi entropies in $2$D.
For a strong-bond cut associated with special surface criticality at the $(2+1)$D $O(3)$ Wilson--Fisher critical point, free-form learning outperforms the calibrated LBW ansatz in predicting the second Rényi entropy and reveals spatially nonuniform coupling renormalizations.
Importantly, the learned EH also provides access to the von Neumann entropy in this setting, revealing a negative subleading correction and extending previous R\'enyi entropy findings to the von Neumann case.
Directly evaluating the von Neumann entropy in such $(2+1)$D interacting quantum many-body systems remains challenging for conventional numerical methods.
These results establish observable-based learning as a route to constructing effective entanglement Hamiltonians beyond prescribed spatial profiles in higher-dimensional interacting systems.
\end{abstract}

\maketitle

\emph{Introduction.---}
Quantum information and condensed matter have recently become increasingly cross-fertilizing~\cite{Amico2008entanglement,laflorencie2016quantum}. 
Quantum entanglement captures many key features of quantum many-body systems, such as topology, criticality, and symmetry breaking~\cite{vidal2003entanglement,Korepin2004universality,Kitaev2006,Levin2006,metlitski2011entanglement}. 
Entanglement entropy and entanglement spectrum have become standard tools for diagnosing these phenomena~\cite{Li2008entangle,Calabrese2008entangle,Fradkin2006entangle,Nussinov2006,Nussinov2009,CASINI2007,JiPRR2019,ji2019categorical,kong2020algebraic,XCWu2020,JRZhao2020,XCWu2021,Thomale2010nonlocal,Poilblanc2010entanglement,schliemann2012entanglement,laflorencie2016quantum,mao2026detecting}. 
Both are encoded in a single object: the entanglement Hamiltonian (EH), defined by $H_A:=-\ln\rho_A$, where $\rho_A=e^{-H_A}$ is the normalized reduced density matrix (RDM)~\cite{Li2008entangle,li2024relevant}.
The von Neumann entropy is thus given by the expectation value of the EH, $S_{\rm vN}=\langle H_A\rangle$, while the entanglement spectrum consists of its eigenvalues. The EH contains all information of the bipartite entanglement, extracting the EH in operator form is therefore a central goal.

Determining the EH of generic interacting systems remains challenging. 
Exact analytical constructions are restricted to special cases such as free-fermionic and integrable systems~\cite{peschel1999density,peschel2003calculation,peschel2009reduced,igloi2010reduced,eisler2017analytical,eisler2018properties}. 
The Bisognano--Wichmann (BW) theorem and its lattice extension (LBW) provide a complementary route~\cite{bisognano1975duality,bisognano1976duality,cardy2016entanglement,dalmonte2018quantum,giudici2018entanglement,mendes2019entanglement,mendes2020measuring,dalmonte2022entanglement}, but they require a nonuniversal velocity scale and assume translational symmetry, and their prescribed spatial profile loses efficacy for inhomogeneous lattices or entanglement-boundary effects~\cite{giudici2018entanglement,yang2026limits,zhu2026bipartite,zhu2026criticalityrenyidefects21d,wang2026universal}. 
Numerical approaches have been explored~\cite{mao2025sampling,Grover2011Entanglement,parisen2018entanglement,zhu2019reconstructing,bairey2019learning,kokail2021entanglement,kokail2021quantum}, among which a density matrix renormalization group (DMRG)-based method has optimized EH couplings by minimizing a Green-function distance from the truncated RDM~\cite{pourjafarabadi2021entanglement}. 
However, this approach requires the RDM information encoded in the Green-function matrix, and its accuracy depends on the bond dimension. Achieving sufficient accuracy may require large bond dimensions, particularly for critical systems and geometries in two or higher spatial dimensions.
These limitations highlight a key challenge: Can we develop a scalable framework for learning effective EHs directly from local observables without reconstructing the RDM?

Here we address this challenge by developing a gradient-based observable-matching framework, using quantum Monte Carlo (QMC) to evaluate the required observables and their responses to Hamiltonian parameters. Since the underlying sampling distribution depends on these parameters, derivatives of expectation values must account for this dependence rather than simply differentiating individual samples.
Inspired by stochastic optimization in variational Monte Carlo, we instead evaluate these derivatives using fluctuation-response relations estimated from Monte Carlo samples~\cite{Harju1997StochasticGradient,Sorella2005WaveFunctionOptimization,Toulouse2007Optimization,Carleo2017NeuralNetworkQuantum}. 
In stochastic series expansion (SSE) QMC, the integrated response of an observable to a coupling is represented by the covariance between the observable and the corresponding operator-string occupation number~\cite{sandvik1991quantum,syljuaasen2002quantum,sandvik1999stochastic,sandvik2010computational}. The required response functions can therefore be estimated from the same QMC samples used to measure the observables, providing stochastic gradients for iterative optimization of the EH couplings. This formulation enables observable-based EH learning within large-scale QMC simulations, offering a route to studying interacting systems in higher spatial dimensions.

\begin{figure*}[t]
    \centering
    \includegraphics[width=\textwidth]{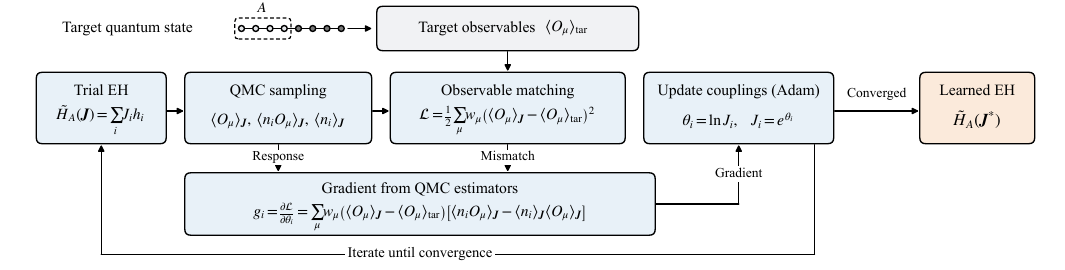}
    \caption{
    Schematic of the EH learning framework.
    Target observables are measured in the original system, while QMC sampling of the trial EH provides the corresponding model observables and their responses to the variational couplings.
    In SSE, the responses are evaluated using covariances between observables and operator counts.
    For the logarithmic parametrization illustrated here, the couplings are updated in $\theta_i=\ln J_i$ using Adam.
    The observable mismatch and responses determine the loss gradient, which is used to iteratively update the variational couplings until convergence.
    }
    \label{fig:learning_framework}
\end{figure*}

We first benchmark coupling recovery and LBW prefactor calibration, then test whether free-form EHs learned from local correlations can predict entropies not used in training. 
The learned EHs closely reproduce independent second R\'enyi entropy benchmarks in $(1+1)$D and $(2+1)$D, as well as von Neumann entropy benchmarks in one dimension. 
In the N\'eel ordered phase of a $(2+1)$D system, the resulting von Neumann entropy estimates are consistent with the expected Goldstone scaling.
At the $(2+1)$D $O(3)$ Wilson--Fisher critical point of the columnar dimerized Heisenberg model, a strong-bond cut associated with nonordinary surface criticality provides a more stringent test.
Here, free-form learning reproduces the replica-QMC $S_2$ benchmark, whereas the calibrated LBW ansatz overestimates it. 
The learned couplings exhibit spatially nonuniform departures from the LBW profile. 
Importantly, the learned EH also provides access to the von Neumann entropy in this setting, revealing a negative subleading correction and extending previous R\'enyi entropy findings to the von Neumann case.
Together, these results demonstrate that observable-based EH learning can capture nontrivial entanglement structure in higher-dimensional interacting systems without reconstructing the full RDM.

\emph{Learning framework.---}
We formulate the determination of an effective EH as an observable-matching problem.
Rather than reconstructing the full reduced density matrix, we introduce a trial EH,
\begin{equation}
    \tilde H_A(\bm J)=\sum_i J_i h_i,
    \label{eq:trial_H}
\end{equation}
where $\{h_i\}$ denotes the chosen operator basis within subsystem $A$.
One simple choice, adopted here, is to take $\{h_i\}$ from the local terms of the original Hamiltonian, consistent with the LBW construction, although the framework allows more general operator choices.
The trial ensemble is
$\tilde\rho_A(\bm J)=e^{-\tilde H_A(\bm J)}/\tilde Z_A(\bm J)$
with $\tilde Z_A(\bm J)=\mathrm{Tr}_A e^{-\tilde H_A(\bm J)}$ for normalization,  
and the effective inverse temperature has been absorbed into $\tilde H_A$.
For a linear observable $O_\mu$ supported in $A$, the target expectation value
$\langle O_\mu\rangle_{\rm tar}=\mathrm{Tr}_A(\rho_A O_\mu)$
can be measured directly in the original system
without reconstructing $\rho_A$.

The corresponding expectation value in the trial ensemble is
$\langle O_\mu\rangle_{\bm J}=\mathrm{Tr}_A[\tilde\rho_A(\bm J)O_\mu]$.
We quantify the mismatch between the trial and target observables through
\begin{equation}
    \mathcal L(\bm J)=\frac12\sum_\mu w_\mu
    \left(
        \langle O_\mu\rangle_{\bm J}
        -\langle O_\mu\rangle_{\rm tar}
    \right)^2,
    \label{eq:loss}
\end{equation}
where $w_\mu$ controls the relative weight of each observable.
The loss gradient is 
\begin{equation}
    \frac{\partial\mathcal L}{\partial J_i}
    =
    \sum_\mu w_\mu
    \left(
        \langle O_\mu\rangle_{\bm J}
        -\langle O_\mu\rangle_{\rm tar}
    \right)
    \frac{\partial\langle O_\mu\rangle_{\bm J}}{\partial J_i}.
    \label{eq:loss_gradient}
\end{equation}
The key quantity required for optimization is therefore the observable response
$\partial\langle O_\mu\rangle_{\bm J}/\partial J_i$.

For observables with no explicit dependence on $J_i$, this response is
given by the imaginary-time fluctuation relation~\cite{kubo1957statistical},
\begin{equation}
    \frac{\partial\langle O_\mu\rangle_{\bm J}}{\partial J_i}
    =
    -\int_0^1 d\tau\,
    \left[
        \langle h_i(\tau)O_\mu\rangle_{\bm J}
        -
        \langle h_i\rangle_{\bm J}
        \langle O_\mu\rangle_{\bm J}
    \right],
    \label{eq:kubo_response}
\end{equation}
where $h_i(\tau)=e^{\tau\tilde H_A}h_i e^{-\tau\tilde H_A}$.
The integration range is $[0,1]$ because the effective inverse temperature has been absorbed into the dimensionless EH.


In the SSE representation, the integrated response in Eq.~\eqref{eq:kubo_response} can be evaluated from covariances between observables and operator-string occupation numbers~\cite{sandvik1999stochastic,sandvik2010computational}.
Let $n_i$ denote the number of nonidentity operators associated with $J_i$ in the SSE operator string.
For observables with coupling-independent SSE estimators, the response becomes
\begin{equation}
    \frac{\partial\langle O_\mu\rangle_{\bm J}}{\partial J_i}
    =
    \frac{
        \langle n_i O_\mu\rangle_{\bm J}
        -
        \langle n_i\rangle_{\bm J}
        \langle O_\mu\rangle_{\bm J}
    }{J_i}.
    \label{eq:sse_response}
\end{equation}
The required response elements can therefore be estimated together with the observables in the same QMC simulation.
Combining these estimates with Eq.~\eqref{eq:loss_gradient} yields the loss gradient directly from equilibrium samples, without differentiating through the QMC sampling procedure.

The couplings are optimized iteratively using the Adam optimizer~\cite{kingma2014adam,bottou2018optimization}.
For positive couplings spanning different magnitudes, we adopt the logarithmic parametrization $\theta_i=\ln J_i$, which allows multiplicative coupling updates while preserving $J_i>0$.
The corresponding response simplifies to
$\partial\langle O_\mu\rangle_{\bm J}/\partial\theta_i
=\langle n_iO_\mu\rangle_{\bm J}
-\langle n_i\rangle_{\bm J}\langle O_\mu\rangle_{\bm J}$.
The learning loop therefore consists of repeated QMC sampling, evaluation of the observable responses, and parameter updates until convergence, as summarized in Fig.~\ref{fig:learning_framework}.
Detailed derivations and implementation procedures are provided in the Supplemental Material.


We first validate the learning procedure by recovering prescribed couplings in single- and multi-parameter benchmarks (see Supplemental Material).
We further demonstrate that equal-time correlations suffice to calibrate the LBW prefactor $2\pi/v_{\mathrm{eff}}$.
The resulting effective velocities approach known bulk values in representative geometries, with finite-size and geometry-dependent deviations (see End Matter).
We next extend the framework from calibrating a single LBW prefactor to learning spatially resolved EH couplings without imposing a prescribed spatial profile.

\emph{Free-form learning of entanglement Hamiltonians.---}
We now move beyond the prescribed LBW spatial profile by learning spatially resolved couplings within a nearest-neighbor Heisenberg ansatz,
\begin{equation}
    \tilde H_A(\bm J)
    =
    \sum_{\langle ij\rangle\subset A}
    J_{ij}\,\bm S_i\cdot\bm S_j.
    \label{eq:freeform_eh}
\end{equation}
We consider half-system bipartitions of $1$D Heisenberg chains and $2$D square-lattice Heisenberg systems, with subsystems of length $L$ in $1$D and size $L\times L$ in $2$D.
The couplings are constrained by the spatial symmetries of each geometry, without imposing any prescribed functional dependence on the distance from the entanglement boundary.
The fixed operator basis defines an approximate EH rather than a reconstruction of all terms in the exact EH~\cite{cirac2011entanglement,li2024relevant,liu2026worldline}. 
We optimize these couplings by matching equal-time nearest-neighbor correlations measured in the original system.
For efficient multi-parameter optimization, we adopt a diagonal-response approximation to the full loss gradient, as detailed in the Supplemental Material.

In practice, we observe that spatially varying EH couplings converge at different rates, with those near the entanglement boundary typically stabilizing first in the uniform systems studied.
This trend is qualitatively consistent with the prominent contribution of degrees of freedom near the entangling boundary to many-body entanglement.
Motivated by this observation, we propose a \emph{sequential-freezing strategy}: we initially optimize all couplings and then progressively freeze stabilized couplings to suppress stochastic drift while continuing to optimize the remaining couplings.
Details of the parametrization and optimization are provided in the Supplemental Material.

\begin{figure}[t]
    \centering
    \includegraphics[width=\columnwidth]{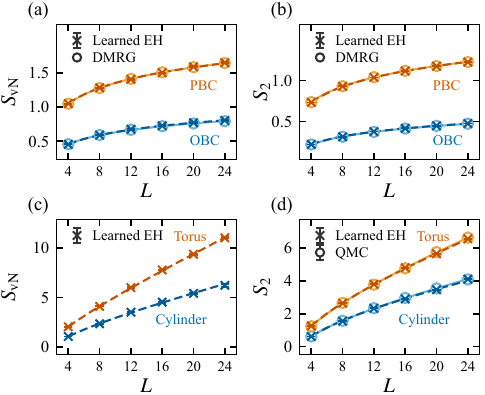}
    \caption{
    Entanglement entropies obtained from free-form EH learning in the spin-$1/2$ Heisenberg model. 
    (a) Von Neumann entropy $S_{\rm vN}$ and (b) second R\'enyi entropy $S_2$ in $1$D for OBC and PBC, compared with DMRG results.
    (c) $S_{\rm vN}$ obtained from the learned EHs for the $2$D cylinder and torus geometries.
    (d) $S_2$ in $2$D for the same geometries, compared with QMC reference results.
    Blue (orange) denotes OBC (PBC) in $1$D and cylinder (torus) geometries in $2$D.
    Crosses denote learned EH results, while open circles denote reference results where available.
    Dashed and solid lines are fits to the learned EH and reference data, respectively, using $a\ln L+b$ in $1$D and $aL+b\ln L+c$ in $2$D.
    Error bars denote Monte Carlo statistical uncertainties.
    }
    \label{fig:freeform_entropies}
\end{figure}

After learning, the von Neumann entropy $S_{\rm vN}$ and the second R\'enyi entropy $S_2$ are evaluated from the thermodynamics of the fixed EH~\cite{mendes2020measuring}.
Unlike the exact EH defined by $H_A=-\ln\rho_A$, the learned EH $\tilde H_A$ does not generally satisfy ${\rm Tr}_A e^{-\tilde H_A}=1$, and its normalization must therefore be explicitly accounted for.
Defining the partition function $\tilde Z_A(\beta_A)={\rm Tr}_A e^{-\beta_A\tilde H_A(\bm J)}$ at effective inverse temperature $\beta_A$, we obtain
\begin{equation}
    \begin{aligned}
        S_{\rm vN}
        &=\ln \tilde Z_A(1)
        +\bigl\langle\tilde H_A(\bm J)\bigr\rangle_{\beta_A=1}, \\ 
        S_2
        &=2\ln \tilde Z_A(1)-\ln \tilde Z_A(2).
    \end{aligned}
    \label{eq:freeform_entropies}
\end{equation}
Note that neither entropy enters the learning objective.
The required partition functions are evaluated using thermal reweight-annealing in SSE~\cite{ding2024reweight}.
Reference entropies are obtained independently from the original systems using DMRG for both entropies in $1$D~\cite{white1992density} and the nonequilibrium replica QMC method for $S_2$ in $2$D~\cite{Hastings2010Measuring,d2020entanglement,zhao2022measuring}, with the latter performed at sufficiently low temperatures to approximate the ground state, as detailed in the Supplemental Material.

For the critical spin-$1/2$ Heisenberg chain, described by a conformal field theory (CFT) with central charge $c=1$, the entanglement entropies exhibit logarithmic scaling. 
For a half-system bipartition, the leading coefficients are $1/6$ and $1/3$ for $S_{\rm vN}$ under open boundary conditions (OBC) and periodic boundary conditions (PBC), respectively, and $1/8$ and $1/4$ for $S_2$, with finite-size corrections in both cases~\cite{holzhey1994geometric,calabrese2004entanglement,calabrese2009entanglement}.
Figs.~\ref{fig:freeform_entropies}(a) and \ref{fig:freeform_entropies}(b) compare the learned entropies with DMRG.
The learned EHs reproduce the logarithmic size dependence of both entropies.
Fits to $a\ln L+b$ yield
$a=0.198(3)$ and $0.330(2)$ for $S_{\rm vN}$ under OBC and PBC, respectively,
and $a=0.1422(4)$ and $0.2757(2)$ for $S_2$.
The learned entropies closely follow the DMRG results, while the fitted coefficients show deviations from the asymptotic CFT predictions over the accessible size range.

In $2$D, the N\'eel state exhibits an area-law contribution and a logarithmic correction from Goldstone modes.  
For the smooth bipartitions considered here, the expected asymptotic form is $aL+b\ln L+c$, with $b=n_G/2=1$ for the two Goldstone modes~\cite{kallin2011anomalies,metlitski2011entanglement,luitz2015universal,kulchytskyy2015detecting,song2024extracting}.
The learned EHs yield the $S_{\rm vN}$ shown in Fig.~\ref{fig:freeform_entropies}(c), with fitted coefficients $b=0.75(9)$ and $0.82(5)$ for the cylinder and torus, respectively.
In Fig.~\ref{fig:freeform_entropies}(d), the learned $S_2$ closely follows the QMC reference for both geometries.
Fits to $aL+b\ln L+c$ yield logarithmic coefficients $b=0.86(2)$ and $1.05(2)$, respectively.
These results support the expected Goldstone-mode logarithmic scaling, with finite-size deviations from the asymptotic prediction $b=1$.

Together, these results demonstrate that effective EHs learned solely from local correlations accurately reproduce independently computed $S_{\rm vN}$ and $S_2$ in $1$D. 
In $2$D, the learned EHs provide estimates of $S_{\rm vN}$ consistent with the expected Goldstone-mode scaling and reproduce independently measured $S_2$ entropies. 
Unlike the LBW-based approach~\cite{mendes2020measuring}, our framework does not require a prescribed spatial profile of the EH couplings.

\begin{figure}[t]
    \centering
    \includegraphics[width=\columnwidth]{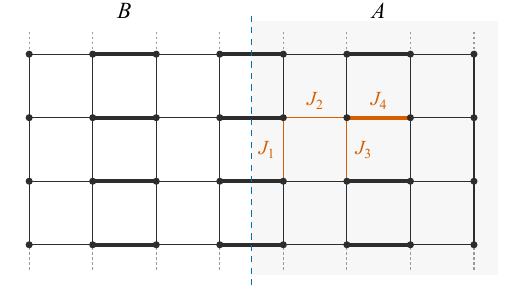}
    \caption{
    Lattice geometry and coupling groups for the strong-bond bipartition of the columnar dimerized Heisenberg model, illustrated for $L=4$.
    The blue dashed line marks the entanglement cut, and the shaded region denotes subsystem $A$.
    Boundary conditions are open along $x$ and periodic along $y$.
    The original Hamiltonian is
    $H=\sum_{\langle ij\rangle}J_{ij}\bm S_i\cdot\bm S_j$,
    where $J_{ij}=J_r$ on the thick bonds and $J_{ij}=1$ on the thin bonds.
    The orange bonds highlight the first four variational coupling groups, $J_1,\ldots,J_4$, with bonds related by translations along $y$ sharing the same coupling.
    }
    \label{fig:sm_qcp_lattice}
\end{figure}

\emph{Learning EHs with boundary effects.---}
We next test free-form learning in a setting where the prescribed LBW profile becomes inaccurate.
We consider the columnar dimerized Heisenberg model at its quantum critical point $J_r=1.90951(1)$~\cite{matsumoto2002ground,ma2018anomalous,zhao2022scaling}, where $J_r$ denotes the ratio of the strong to weak exchange couplings (see Fig.~\ref{fig:sm_qcp_lattice}).
We choose the entanglement cut to pass through the strong bonds.
This bipartition is associated with effective dangling spin-$1/2$ degrees of freedom along the entanglement boundary and with a special surface criticality~\cite{ding2018engineering,weber2018nonordinary,weber2021spin,wang2026universal,zhu2026bipartite,zhu2026criticalityrenyidefects21d}.
The previous work found systematic deviations of LBW entanglement correlations from those of the original system, even after calibrating the overall scale~\cite{yang2026limits}. In other words, there is no good approximation for this case to estimate the EH and other entanglement observables.

We consider a $2L\times L$ cylinder with open boundaries along $x$ and periodic boundaries along $y$.
The horizontal bonds alternate between strong and weak exchanges, while the vertical bonds are weak.
An equal bipartition across strong bonds defines an $L\times L$ subsystem $A$.
We learn the nearest-neighbor EH in Eq.~\eqref{eq:freeform_eh} from equal-time correlations, imposing translation symmetry along $y$ without prescribing the spatial coupling profile, as shown in Fig.~\ref{fig:sm_qcp_lattice}.
For comparison, the LBW ansatz uses the same operator basis but fixes the spatial profile and the prefactor $2\pi/v$, with $v$ determined from imaginary-time correlations~\cite{yang2026limits}.
The learned-EH entropies and the LBW $S_2$ are evaluated thermodynamically on subsystem $A$, while the QMC reference for $S_2$ is obtained independently from the original system using the nonequilibrium replica QMC method described above.

\begin{figure}[t]
    \centering
    \includegraphics[width=0.9\columnwidth]{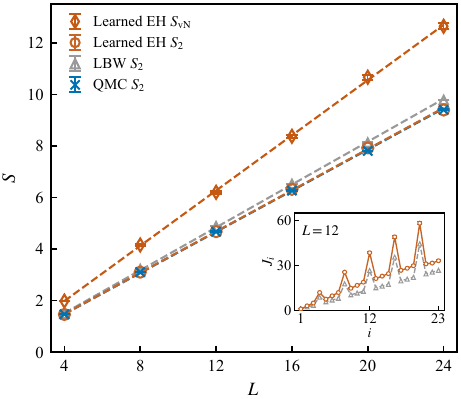}
    \caption{
    Von Neumann entropy $S_{\rm vN}$ and second R\'enyi entropy $S_2$ for an $L\times L$ subsystem of the columnar dimerized Heisenberg model at its quantum critical point $J_r=1.90951(1)$, with an entanglement cut through strong bonds.
    Orange diamonds and circles denote the learned-EH $S_{\rm vN}$ and $S_2$, respectively, while gray triangles and blue crosses denote the LBW and QMC results for $S_2$.
    In the main panel, dashed lines represent weighted fits to $S(L)=aL+c$ for the corresponding datasets.
    The inset compares the learned (orange circles) and LBW (gray triangles) nearest-neighbor couplings at $L=12$, with bond-group index $i$ following the ordering in Fig.~\ref{fig:sm_qcp_lattice} from the entanglement cut into subsystem $A$.
    Error bars denote Monte Carlo statistical uncertainties for the entropies and standard errors across independent learning runs for the learned couplings in the inset.
    }
    \label{fig:qcp_entropy}
\end{figure}

As shown in Fig.~\ref{fig:qcp_entropy}, the learned EH reproduces the QMC benchmark for $S_2$, substantially reducing the systematic overestimation by the LBW ansatz, particularly at larger system sizes.
Remarkably, this agreement is achieved without using any entanglement entropy in the learning process, providing an independent validation of the learned EH.
To quantify the $S_2$ agreement, we fit the learned-EH, QMC reference, and LBW results for $L=4,8,12,16,20,24$ to $S_2(L)=aL+c$.
The learned EH yields $a=0.4000(4)$ and $c=-0.116(4)$, in close agreement with the QMC reference values $a=0.3989(5)$ and $c=-0.117(4)$. In contrast, the calibrated LBW ansatz gives $a=0.4156(4)$ and $c=-0.145(5)$, exhibiting substantially larger deviations. 
These fits quantify the agreement over the accessible system sizes rather than determine the asymptotic scaling.

The improved accuracy is accompanied by nontrivial modifications of the effective EH couplings. 
As shown in the inset of Fig.~\ref{fig:qcp_entropy} for $L=12$, the learned couplings retain the overall spatial pattern of the LBW profile but exhibit nonuniform enhancements extending into the subsystem. 
Such deviations cannot be captured by simply rescaling the overall LBW prefactor, indicating that the boundary effects cannot be fully captured by a simple rescaling of the LBW profile.


Beyond the independently benchmarked $S_2$, the learned EH provides access to $S_{\text{vN}}$ at this interacting quantum critical point, a quantity that is challenging to obtain directly using conventional replica QMC. 
The subleading correction of $S_{\text{vN}}$ at the Wilson--Fisher fixed point has been investigated theoretically by previous studies using $\epsilon$- and large-$N$ expansions~\cite{metlitski2009entanglement,whitsitt2017entanglement}. However, these approaches do not establish the sign of the subleading correction for the present strong-bond-cut geometry with periodic boundary conditions.
Remarkably, our learned EH yields a negative subleading correction to $S_{\rm vN}$ beyond the leading area law. A weighted fit to $S_{\rm vN}(L)=a_{\rm vN}L+c_{\rm vN}$ yields $a_{\rm vN}=0.536(3)$ and $c_{\rm vN}=-0.16(2)$. 
This negative offset agrees in sign with the $S_2$ correction found in the present geometry and in previous strong-bond-cut studies~\cite{wang2026universal,wang2026boundary},
extending the numerical investigation of nontrivial boundary entanglement to the von Neumann case.



These results demonstrate that observable-based EH learning captures nontrivial entanglement structure beyond the prescribed LBW profile, reproducing independently benchmarked $S_2$ substantially more accurately than the calibrated LBW ansatz, 
while also resolving a negative subleading offset in the learned EH $S_{\rm vN}$ over the accessible sizes and revealing spatially nonuniform boundary-induced modifications of the effective couplings.

\emph{Conclusion and outlook.---}
We have developed an observable-based framework for learning effective entanglement Hamiltonians (EHs) from local correlations and demonstrated its implementation in SSE QMC using fluctuation-response estimators. After validating the framework by recovering known couplings and calibrating the overall prefactor of the LBW construction, we extend it to free-form learning of spatially resolved EH couplings. Without using entropy data in training, the learned EHs accurately reproduce reference second R\'enyi entropies in $1$D and $2$D and von Neumann entropies in $1$D. In the N\'eel-ordered $2$D systems, they further yield von Neumann entropy estimates consistent with the expected Goldstone scaling.
At the $(2+1)$D $O(3)$ Wilson--Fisher quantum critical point of the columnar dimerized Heisenberg model, free-form learning substantially improves the second R\'enyi entropy prediction over the calibrated LBW ansatz for a special entanglement-cut, while revealing spatially nonuniform boundary-induced modifications of the effective couplings. 
Importantly, the learned EH also provides access to the von Neumann entropy in this setting, revealing a negative subleading correction and extending previous R\'enyi entropy findings to the von Neumann case.
These results demonstrate that learning effective EHs from local observables offers a powerful route to probing nontrivial entanglement structure in higher-dimensional interacting systems without reconstructing the full reduced density matrix.
A natural next step is to systematically enlarge the operator basis and the set of target observables, including long-range, anisotropic, and multispin terms~\cite{cirac2011entanglement,li2024relevant,liu2026worldline}. This would allow the learned EH to resolve operator structures induced by lattice and entanglement-boundary effects beyond an effective renormalization of local couplings. More broadly, an explicit learned EH provides access to thermodynamic and spectral properties, opening new possibilities for studying entanglement beyond integer-order R\'enyi entropies.

\emph{Acknowledgments.---}
SY and YL contribute equally to this work.
This project is supported by the Scientific Research Project (No. WU2025B011), Feng-Ying Career Development Chair Fund and the Start-up Funding of Westlake University, and the Natural Science Foundation of China (Grant No. 12674188), and Quantum MetaKey (QuaMeK) computing. 
The authors thank the IT service office and the high-performance computing center of Westlake University.

\bibliography{ref}

@article{laflorencie2016quantum,
  title={Quantum entanglement in condensed matter systems},
  author={Laflorencie, Nicolas},
  journal={Physics Reports},
  volume={646},
  pages={1--59},
  year={2016},
  doi={10.1016/j.physrep.2016.06.008},
  url={https://doi.org/10.1016/j.physrep.2016.06.008}
}

@article{mao2025sampling,
  title={Sampling reduced density matrix to extract fine levels of entanglement spectrum and restore entanglement Hamiltonian},
  author={Mao, Bin-Bin and Ding, Yi-Ming and Wang, Zhe and Hu, Shijie and Yan, Zheng},
  journal={Nature Communications},
  volume={16},
  number={1},
  pages={2880},
  year={2025},
  doi={10.1038/s41467-025-56867-9},
  url={https://doi.org/10.1038/s41467-025-56867-9}
}

@article{mao2026detecting,
  title={Detecting the emergent continuous symmetry of criticality via a subsystem’s entanglement spectrum},
  author={Mao, Bin-Bin and Wang, Zhe and Chen, Bin-Bin and Yan, Zheng},
  journal={Physical Review Letters},
  volume={136},
  number={4},
  pages={046401},
  year={2026},
  doi={10.1103/PhysRevLett.136.046401},
  url={https://doi.org/10.1103/PhysRevLett.136.046401}
}

@article{peschel1999density,
  title={Density-matrix spectra for integrable models},
  author={Peschel, Ingo and Kaulke, Matthias and Legeza, {\"O}rs},
  journal={Annalen der Physik},
  volume={511},
  number={2},
  pages={153--164},
  year={1999},
  doi={10.1002/(SICI)1521-3889(199911)511:2<153::AID-ANDP153>3.0.CO;2-2},
  url={https://doi.org/10.1002/(SICI)1521-3889(199911)511:2<153::AID-ANDP153>3.0.CO;2-2}
}

@article{peschel2003calculation,
  title={Calculation of reduced density matrices from correlation functions},
  author={Peschel, Ingo},
  journal={Journal of Physics A: Mathematical and General},
  volume={36},
  number={14},
  pages={L205--L208},
  year={2003},
  doi={10.1088/0305-4470/36/14/101},
  url={https://doi.org/10.1088/0305-4470/36/14/101}
}

@article{peschel2009reduced,
  title={Reduced density matrices and entanglement entropy in free lattice models},
  author={Peschel, Ingo and Eisler, Viktor},
  journal={Journal of Physics A: Mathematical and Theoretical},
  volume={42},
  number={50},
  pages={504003},
  year={2009},
  doi={10.1088/1751-8113/42/50/504003},
  url={https://doi.org/10.1088/1751-8113/42/50/504003}
}

@article{igloi2010reduced,
  title={On reduced density matrices for disjoint subsystems},
  author={Igl{\'o}i, Ferenc and Peschel, Ingo},
  journal={EPL (Europhysics Letters)},
  volume={89},
  number={4},
  pages={40001},
  year={2010},
  doi={10.1209/0295-5075/89/40001},
  url={https://doi.org/10.1209/0295-5075/89/40001}
}

@article{eisler2017analytical,
  title={Analytical results for the entanglement Hamiltonian of a free-fermion chain},
  author={Eisler, Viktor and Peschel, Ingo},
  journal={Journal of Physics A: Mathematical and Theoretical},
  volume={50},
  number={28},
  pages={284003},
  year={2017},
  doi={10.1088/1751-8121/aa7498},
  url={https://doi.org/10.1088/1751-8121/aa7498}
}

@article{eisler2018properties,
  title={Properties of the entanglement Hamiltonian for finite free-fermion chains},
  author={Eisler, Viktor and Peschel, Ingo},
  journal={Journal of Statistical Mechanics: Theory and Experiment},
  volume={2018},
  number={10},
  pages={104001},
  year={2018},
  doi={10.1088/1742-5468/aae2e4},
  url={https://doi.org/10.1088/1742-5468/aae2e4}
}

@article{bisognano1975duality,
  title={On the duality condition for a Hermitian scalar field},
  author={Bisognano, Joseph J and Wichmann, Eyvind H},
  journal={Journal of Mathematical Physics},
  volume={16},
  number={4},
  pages={985--1007},
  year={1975},
  doi={10.1063/1.522605},
  url={https://doi.org/10.1063/1.522605}
}

@article{bisognano1976duality,
  title={On the duality condition for quantum fields},
  author={Bisognano, Joseph J and Wichmann, Eyvind H},
  journal={Journal of Mathematical Physics},
  volume={17},
  number={3},
  pages={303--321},
  year={1976},
  doi={10.1063/1.522898},
  url={https://doi.org/10.1063/1.522898}
}

@article{dalmonte2018quantum,
  title={Quantum simulation and spectroscopy of entanglement Hamiltonians},
  author={Dalmonte, Marcello and Vermersch, Beno{\^\i}t and Zoller, Peter},
  journal={Nature Physics},
  volume={14},
  number={8},
  pages={827--831},
  year={2018},
  doi={10.1038/s41567-018-0151-7},
  url={https://doi.org/10.1038/s41567-018-0151-7}
}

@article{giudici2018entanglement,
  title={Entanglement Hamiltonians of lattice models via the Bisognano-Wichmann theorem},
  author={Giudici, G and Mendes-Santos, Tiago and Calabrese, P and Dalmonte, Marcello},
  journal={Physical Review B},
  volume={98},
  number={13},
  pages={134403},
  year={2018},
  doi={10.1103/PhysRevB.98.134403},
  url={https://doi.org/10.1103/PhysRevB.98.134403}
}

@article{mendes2019entanglement,
  title={Entanglement Hamiltonian of quantum critical chains and conformal field theories},
  author={Mendes-Santos, Tiago and Giudici, G and Dalmonte, M and Rajabpour, MA},
  journal={Physical Review B},
  volume={100},
  number={15},
  pages={155122},
  year={2019},
  doi={10.1103/PhysRevB.100.155122},
  url={https://doi.org/10.1103/PhysRevB.100.155122}
}

@article{mendes2020measuring,
  title={Measuring von Neumann entanglement entropies without wave functions},
  author={Mendes-Santos, Tiago and Giudici, G and Fazio, R and Dalmonte, M},
  journal={New Journal of Physics},
  volume={22},
  number={1},
  pages={013044},
  year={2020},
  doi={10.1088/1367-2630/ab6875},
  url={https://doi.org/10.1088/1367-2630/ab6875}
}

@article{dalmonte2022entanglement,
  title={Entanglement Hamiltonians: from field theory to lattice models and experiments},
  author={Dalmonte, Marcello and Eisler, Viktor and Falconi, Marco and Vermersch, Beno{\^\i}t},
  journal={Annalen der Physik},
  volume={534},
  number={11},
  pages={2200064},
  year={2022},
  doi={10.1002/andp.202200064},
  url={https://doi.org/10.1002/andp.202200064}
}

@article{yang2026limits,
  title={Limits of the lattice Bisognano-Wichmann form for entanglement Hamiltonians: A quantum Monte Carlo study},
  author={Yang, Siyi and Ding, Yi-Ming and Yan, Zheng},
  journal={Physical Review B},
  volume={113},
  number={23},
  pages={235115},
  year={2026},
  doi={10.1103/PhysRevB.113.235115},
  url={https://doi.org/10.1103/PhysRevB.113.235115}
}

@article{kubo1957statistical,
  title={Statistical-mechanical theory of irreversible processes. I. General theory and simple applications to magnetic and conduction problems},
  author={Kubo, Ryogo},
  journal={Journal of the Physical Society of Japan},
  volume={12},
  number={6},
  pages={570--586},
  year={1957},
  doi={10.1143/JPSJ.12.570},
  url={https://doi.org/10.1143/JPSJ.12.570}
}

@article{sandvik1999stochastic,
  title={Stochastic series expansion method with operator-loop update},
  author={Sandvik, Anders W},
  journal={Physical Review B},
  volume={59},
  number={22},
  pages={R14157},
  year={1999},
  doi={10.1103/PhysRevB.59.R14157},
  url={https://doi.org/10.1103/PhysRevB.59.R14157}
}

@inproceedings{kingma2014adam,
  title     = {{Adam}: A Method for Stochastic Optimization},
  author    = {Kingma, Diederik P. and Ba, Jimmy},
  booktitle = {International Conference on Learning Representations (ICLR)},
  year      = {2015},
  url       = {https://arxiv.org/abs/1412.6980}
}

@article{sen2015velocity,
  title={Velocity of excitations in ordered, disordered, and critical antiferromagnets},
  author={Sen, Arnab and Suwa, Hidemaro and Sandvik, Anders W},
  journal={Physical Review B},
  volume={92},
  number={19},
  pages={195145},
  year={2015},
  doi={10.1103/PhysRevB.92.195145},
  url={https://doi.org/10.1103/PhysRevB.92.195145}
}

@article{des1962spin,
  title={Spin-wave spectrum of the antiferromagnetic linear chain},
  author={Des Cloizeaux, Jacques and Pearson, JJ},
  journal={Physical Review},
  volume={128},
  number={5},
  pages={2131},
  year={1962},
  doi={10.1103/PhysRev.128.2131},
  url={https://doi.org/10.1103/PhysRev.128.2131}
}

@article{beard1998square,
  title={Square-lattice Heisenberg antiferromagnet at very large correlation lengths},
  author={Beard, BB and Birgeneau, RJ and Greven, M and Wiese, U-J},
  journal={Physical Review Letters},
  volume={80},
  number={8},
  pages={1742},
  year={1998},
  doi={10.1103/PhysRevLett.80.1742},
  url={https://doi.org/10.1103/PhysRevLett.80.1742}
}

@article{ma2018anomalous,
  title={Anomalous quantum-critical scaling corrections in two-dimensional antiferromagnets},
  author={Ma, Nvsen and Weinberg, Phillip and Shao, Hui and Guo, Wenan and Yao, Dao-Xin and Sandvik, Anders W},
  journal={Physical Review Letters},
  volume={121},
  number={11},
  pages={117202},
  year={2018},
  doi={10.1103/PhysRevLett.121.117202},
  url={https://doi.org/10.1103/PhysRevLett.121.117202}
}

@article{sandvik1991quantum,
  title={Quantum Monte Carlo simulation method for spin systems},
  author={Sandvik, Anders W and Kurkij{\"a}rvi, Juhani},
  journal={Physical Review B},
  volume={43},
  number={7},
  pages={5950},
  year={1991},
  doi={10.1103/PhysRevB.43.5950},
  url={https://doi.org/10.1103/PhysRevB.43.5950}
}

@article{syljuaasen2002quantum,
  title={Quantum Monte Carlo with directed loops},
  author={Sylju{\aa}sen, Olav F and Sandvik, Anders W},
  journal={Physical Review E},
  volume={66},
  number={4},
  pages={046701},
  year={2002},
  doi={10.1103/PhysRevE.66.046701},
  url={https://doi.org/10.1103/PhysRevE.66.046701}
}

@inproceedings{sandvik2010computational,
  title={Computational studies of quantum spin systems},
  author={Sandvik, Anders W},
  booktitle={AIP Conference Proceedings},
  volume={1297},
  number={1},
  pages={135--338},
  year={2010},
  doi={10.1063/1.3518900},
  url={https://doi.org/10.1063/1.3518900}
}

@article{metlitski2011entanglement,
  title={Entanglement entropy of systems with spontaneously broken continuous symmetry},
  author={Metlitski, Max A and Grover, Tarun},
  journal={arXiv preprint arXiv:1112.5166},
  year={2011},
  url={https://arxiv.org/abs/1112.5166}
}

@article{calabrese2009entanglement,
  title={Entanglement entropy and conformal field theory},
  author={Calabrese, Pasquale and Cardy, John},
  journal={Journal of Physics A: Mathematical and Theoretical},
  volume={42},
  number={50},
  pages={504005},
  year={2009},
  doi={10.1088/1751-8113/42/50/504005},
  url={https://doi.org/10.1088/1751-8113/42/50/504005}
}

@article{ding2024reweight,
  title={Reweight-annealing method for evaluating the partition function via quantum Monte Carlo calculations},
  author={Ding, Yi-Ming and Sun, Jun-Song and Ma, Nvsen and Pan, Gaopei and Cheng, Chen and Yan, Zheng},
  journal={Physical Review B},
  volume={110},
  number={16},
  pages={165152},
  year={2024},
  doi={10.1103/PhysRevB.110.165152},
  url={https://doi.org/10.1103/PhysRevB.110.165152}
}

@article{d2020entanglement,
  title={Entanglement entropy from nonequilibrium work},
  author={D’Emidio, Jonathan},
  journal={Physical Review Letters},
  volume={124},
  number={11},
  pages={110602},
  year={2020},
  doi={10.1103/PhysRevLett.124.110602},
  url={https://doi.org/10.1103/PhysRevLett.124.110602}
}

@article{zhao2022measuring,
  title={Measuring R{\'e}nyi entanglement entropy with high efficiency and precision in quantum Monte Carlo simulations},
  author={Zhao, Jiarui and Chen, Bin-Bin and Wang, Yan-Cheng and Yan, Zheng and Cheng, Meng and Meng, Zi Yang},
  journal={npj Quantum Materials},
  volume={7},
  number={1},
  pages={69},
  year={2022},
  doi={10.1038/s41535-022-00476-0},
  url={https://doi.org/10.1038/s41535-022-00476-0}
}

@article{zhu2026bipartite,
  title={Bipartite entanglement and surface criticality: The extra contribution of the nonordinary edge in entanglement},
  author={Zhu, Yanzhang and Liu, Zenan and Wang, Zhe and Wang, Yan-Cheng and Yan, Zheng},
  journal={Physical Review Letters},
  volume={136},
  number={4},
  pages={046501},
  year={2026},
  doi={10.1103/PhysRevLett.136.046501},
  url={https://doi.org/10.1103/PhysRevLett.136.046501}
}

@article{wang2026universal,
  title={Universal and non-universal contributions of entanglement under different bipartitions},
  author={Wang, Zhe and Guo, Chunhao and Mao, Bin-Bin and Yan, Zheng},
  journal={arXiv preprint arXiv:2601.12365},
  year={2026},
  url={https://arxiv.org/abs/2601.12365}
}

@article{li2024relevant,
  title={Relevant long-range interaction of the entanglement hamiltonian emerges from a short-range gapped system},
  author={Li, Chuhao and Huang, Rui-Zhen and Ding, Yi-Ming and Meng, Zi Yang and Wang, Yan-Cheng and Yan, Zheng},
  journal={Physical Review B},
  volume={109},
  number={19},
  pages={195169},
  year={2024},
  doi={10.1103/PhysRevB.109.195169},
  url={https://doi.org/10.1103/PhysRevB.109.195169}
}

@article{liu2026worldline,
  title={Worldline deconfinement and emergent long-range interaction in the entanglement Hamiltonian and in the entanglement spectrum},
  author={Liu, Zenan and Wang, Zhe and Yao, Dao-Xin and Yan, Zheng},
  journal={Physical Review B},
  volume={113},
  number={14},
  pages={144425},
  year={2026},
  doi={10.1103/PhysRevB.113.144425},
  url={https://doi.org/10.1103/PhysRevB.113.144425}
}

@article{bairey2019learning,
  title={Learning a local Hamiltonian from local measurements},
  author={Bairey, Eyal and Arad, Itai and Lindner, Netanel H},
  journal={Physical Review Letters},
  volume={122},
  number={2},
  pages={020504},
  year={2019},
  doi={10.1103/PhysRevLett.122.020504},
  url={https://doi.org/10.1103/PhysRevLett.122.020504}
}

@article{kokail2021entanglement,
  title={Entanglement Hamiltonian tomography in quantum simulation},
  author={Kokail, Christian and van Bijnen, Rick and Elben, Andreas and Vermersch, Beno{\^\i}t and Zoller, Peter},
  journal={Nature Physics},
  volume={17},
  number={8},
  pages={936--942},
  year={2021},
  doi={10.1038/s41567-021-01260-w},
  url={https://doi.org/10.1038/s41567-021-01260-w}
}

@article{kokail2021quantum,
  title={Quantum variational learning of the entanglement Hamiltonian},
  author={Kokail, Christian and Sundar, Bhuvanesh and Zache, Torsten V and Elben, Andreas and Vermersch, Beno{\^\i}t and Dalmonte, Marcello and van Bijnen, Rick and Zoller, Peter},
  journal={Physical Review Letters},
  volume={127},
  number={17},
  pages={170501},
  year={2021},
  doi={10.1103/PhysRevLett.127.170501},
  url={https://doi.org/10.1103/PhysRevLett.127.170501}
}

@article{bottou2018optimization,
  title={Optimization methods for large-scale machine learning},
  author={Bottou, L{\'e}on and Curtis, Frank E and Nocedal, Jorge},
  journal={SIAM Review},
  volume={60},
  number={2},
  pages={223--311},
  year={2018},
  doi={10.1137/16M1080173},
  url={https://doi.org/10.1137/16M1080173}
}

@article{pourjafarabadi2021entanglement,
  title={Entanglement Hamiltonian of interacting systems: Local temperature approximation and beyond},
  author={Pourjafarabadi, Mahdieh and Najafzadeh, Hanieh and Vaezi, Mohammad-Sadegh and Vaezi, Abolhassan},
  journal={Physical Review Research},
  volume={3},
  number={1},
  pages={013217},
  year={2021},
  doi={10.1103/PhysRevResearch.3.013217},
  url={https://doi.org/10.1103/PhysRevResearch.3.013217}
}

@article{white1992density,
  title={Density matrix formulation for quantum renormalization groups},
  author={White, Steven R},
  journal={Physical Review Letters},
  volume={69},
  number={19},
  pages={2863},
  year={1992},
  doi={10.1103/PhysRevLett.69.2863},
  url={https://doi.org/10.1103/PhysRevLett.69.2863}
}

@article{parisen2018entanglement,
  title={Entanglement Hamiltonian of interacting fermionic models},
  author={Parisen Toldin, Francesco and Assaad, Fakher F},
  journal={Physical Review Letters},
  volume={121},
  number={20},
  pages={200602},
  year={2018},
  doi={10.1103/PhysRevLett.121.200602},
  url={https://doi.org/10.1103/PhysRevLett.121.200602}
}

@article{zhu2019reconstructing,
  title={Reconstructing entanglement Hamiltonian via entanglement eigenstates},
  author={Zhu, Wei and Huang, Zhoushen and He, Yin-Chen},
  journal={Physical Review B},
  volume={99},
  number={23},
  pages={235109},
  year={2019},
  doi={10.1103/PhysRevB.99.235109},
  url={https://doi.org/10.1103/PhysRevB.99.235109}
}

@article{cardy2016entanglement,
  title={Entanglement Hamiltonians in two-dimensional conformal field theory},
  author={Cardy, John and Tonni, Erik},
  journal={Journal of Statistical Mechanics: Theory and Experiment},
  volume={2016},
  number={12},
  pages={123103},
  year={2016},
  doi={10.1088/1742-5468/2016/12/123103},
  url={https://doi.org/10.1088/1742-5468/2016/12/123103}
}

@article{ding2018engineering,
  title={Engineering surface critical behavior of (2+1)-dimensional O(3) quantum critical points},
  author={Ding, Chengxiang and Zhang, Long and Guo, Wenan},
  journal={Physical Review Letters},
  volume={120},
  number={23},
  pages={235701},
  year={2018},
  doi={10.1103/PhysRevLett.120.235701},
  url={https://doi.org/10.1103/PhysRevLett.120.235701}
}

@article{schliemann2012entanglement,
  title={Entanglement spectra of Heisenberg ladders of higher spin},
  author={Schliemann, John and L{\"a}uchli, Andreas M},
  journal={Journal of Statistical Mechanics: Theory and Experiment},
  volume={2012},
  number={11},
  pages={P11021},
  year={2012},
  doi={10.1088/1742-5468/2012/11/P11021},
  url={https://doi.org/10.1088/1742-5468/2012/11/P11021}
}

@article{cirac2011entanglement,
  title={Entanglement spectrum and boundary theories with projected entangled-pair states},
  author={Cirac, J Ignacio and Poilblanc, Didier and Schuch, Norbert and Verstraete, Frank},
  journal={Physical Review B},
  volume={83},
  number={24},
  pages={245134},
  year={2011},
  doi={10.1103/PhysRevB.83.245134},
  url={https://doi.org/10.1103/PhysRevB.83.245134}
}

@article{Amico2008entanglement,
  title={Entanglement in many-body systems},
  author={Amico, Luigi and Fazio, Rosario and Osterloh, Andreas and Vedral, Vlatko},
  journal={Reviews of Modern Physics},
  volume={80},
  number={2},
  pages={517--576},
  year={2008},
  doi={10.1103/RevModPhys.80.517},
  url={https://doi.org/10.1103/RevModPhys.80.517}
}

@article{vidal2003entanglement,
  title={Entanglement in quantum critical phenomena},
  author={Vidal, G and Latorre, J I and Rico, E and Kitaev, A},
  journal={Physical Review Letters},
  volume={90},
  number={22},
  pages={227902},
  year={2003},
  doi={10.1103/PhysRevLett.90.227902},
  url={https://doi.org/10.1103/PhysRevLett.90.227902}
}

@article{Poilblanc2010entanglement,
  title={Entanglement spectra of quantum Heisenberg ladders},
  author={Poilblanc, Didier},
  journal={Physical Review Letters},
  volume={105},
  number={7},
  pages={077202},
  year={2010},
  doi={10.1103/PhysRevLett.105.077202},
  url={https://doi.org/10.1103/PhysRevLett.105.077202}
}

@article{Li2008entangle,
  title={Entanglement spectrum as a generalization of entanglement entropy: Identification of topological order in non-Abelian fractional quantum Hall effect states},
  author={Li, Hui and Haldane, F D M},
  journal={Physical Review Letters},
  volume={101},
  number={1},
  pages={010504},
  year={2008},
  doi={10.1103/PhysRevLett.101.010504},
  url={https://doi.org/10.1103/PhysRevLett.101.010504}
}

@article{Calabrese2008entangle,
  title={Entanglement spectrum in one-dimensional systems},
  author={Calabrese, Pasquale and Lefevre, Alexandre},
  journal={Physical Review A},
  volume={78},
  number={3},
  pages={032329},
  year={2008},
  doi={10.1103/PhysRevA.78.032329},
  url={https://doi.org/10.1103/PhysRevA.78.032329}
}

@article{Thomale2010nonlocal,
  title={Nonlocal order in gapless systems: Entanglement spectrum in spin chains},
  author={Thomale, Ronny and Arovas, D P and Bernevig, B Andrei},
  journal={Physical Review Letters},
  volume={105},
  number={11},
  pages={116805},
  year={2010},
  doi={10.1103/PhysRevLett.105.116805},
  url={https://doi.org/10.1103/PhysRevLett.105.116805}
}

@article{Korepin2004universality,
  title={Universality of entropy scaling in one dimensional gapless models},
  author={Korepin, V E},
  journal={Physical Review Letters},
  volume={92},
  number={9},
  pages={096402},
  year={2004},
  doi={10.1103/PhysRevLett.92.096402},
  url={https://doi.org/10.1103/PhysRevLett.92.096402}
}

@article{Fradkin2006entangle,
  title={Entanglement entropy of 2D conformal quantum critical points: Hearing the shape of a quantum drum},
  author={Fradkin, Eduardo and Moore, Joel E},
  journal={Physical Review Letters},
  volume={97},
  number={5},
  pages={050404},
  year={2006},
  doi={10.1103/PhysRevLett.97.050404},
  url={https://doi.org/10.1103/PhysRevLett.97.050404}
}

@article{CASINI2007,
  title={Universal terms for the entanglement entropy in 2+1 dimensions},
  author={Casini, H and Huerta, M},
  journal={Nuclear Physics B},
  volume={764},
  number={3},
  pages={183--201},
  year={2007},
  doi={10.1016/j.nuclphysb.2006.12.012},
  url={https://doi.org/10.1016/j.nuclphysb.2006.12.012}
}

@article{zhu2026criticalityrenyidefects21d,
  title = {Criticality on R\'enyi Defects at $(2+1)\mathrm{D}$ O(3) Quantum Critical Points},
  author = {Zhu, Yanzhang and Wang, Zhe and Cheng, Meng and Yan, Zheng},
  journal = {Phys. Rev. Lett.},
  volume = {137},
  issue = {15},
  pages = {156501},
  numpages = {7},
  year = {2026},
  month = {Oct},
  publisher = {American Physical Society},
  doi = {10.1103/7zzj-78fv},
  url = {https://link.aps.org/doi/10.1103/7zzj-78fv}
}

@article{Kitaev2006,
  title={Topological entanglement entropy},
  author={Kitaev, Alexei and Preskill, John},
  journal={Physical Review Letters},
  volume={96},
  number={11},
  pages={110404},
  year={2006},
  doi={10.1103/PhysRevLett.96.110404},
  url={https://doi.org/10.1103/PhysRevLett.96.110404}
}

@article{Levin2006,
  title={Detecting topological order in a ground state wave function},
  author={Levin, Michael and Wen, Xiao-Gang},
  journal={Physical Review Letters},
  volume={96},
  number={11},
  pages={110405},
  year={2006},
  doi={10.1103/PhysRevLett.96.110405},
  url={https://doi.org/10.1103/PhysRevLett.96.110405}
}

@article{JRZhao2020,
  title={Higher-form symmetry breaking at Ising transitions},
  author={Zhao, Jiarui and Yan, Zheng and Cheng, Meng and Meng, Zi Yang},
  journal={Physical Review Research},
  volume={3},
  number={3},
  pages={033024},
  year={2021},
  doi={10.1103/PhysRevResearch.3.033024},
  url={https://doi.org/10.1103/PhysRevResearch.3.033024}
}

@article{JiPRR2019,
  title={Noninvertible anomalies and mapping-class-group transformation of anomalous partition functions},
  author={Ji, Wenjie and Wen, Xiao-Gang},
  journal={Physical Review Research},
  volume={1},
  number={3},
  pages={033054},
  year={2019},
  doi={10.1103/PhysRevResearch.1.033054},
  url={https://doi.org/10.1103/PhysRevResearch.1.033054}
}

@article{ji2019categorical,
  title={Categorical symmetry and noninvertible anomaly in symmetry-breaking and topological phase transitions},
  author={Ji, Wenjie and Wen, Xiao-Gang},
  journal={Physical Review Research},
  volume={2},
  number={3},
  pages={033417},
  year={2020},
  doi={10.1103/PhysRevResearch.2.033417},
  url={https://doi.org/10.1103/PhysRevResearch.2.033417}
}

@article{kong2020algebraic,
  title={Algebraic higher symmetry and categorical symmetry: A holographic and entanglement view of symmetry},
  author={Kong, Liang and Lan, Tian and Wen, Xiao-Gang and Zhang, Zhi-Hao and Zheng, Hao},
  journal={Physical Review Research},
  volume={2},
  number={4},
  pages={043086},
  year={2020},
  doi={10.1103/PhysRevResearch.2.043086},
  url={https://doi.org/10.1103/PhysRevResearch.2.043086}
}

@article{Nussinov2006,
  title={Sufficient symmetry conditions for topological quantum order},
  author={Nussinov, Zohar and Ortiz, Gerardo},
  journal={Proceedings of the National Academy of Sciences},
  volume={106},
  pages={16944--16949},
  year={2009},
  doi={10.1073/pnas.0803726105},
  url={https://doi.org/10.1073/pnas.0803726105}
}

@article{Nussinov2009,
  title={A symmetry principle for topological quantum order},
  author={Nussinov, Zohar and Ortiz, Gerardo},
  journal={Annals of Physics},
  volume={324},
  pages={977--1057},
  year={2009},
  doi={10.1016/j.aop.2008.11.002},
  url={https://doi.org/10.1016/j.aop.2008.11.002}
}

@article{XCWu2020,
  title={Categorical symmetries at criticality},
  author={Wu, Xiao-Chuan and Ji, Wenjie and Xu, Cenke},
  journal={Journal of Statistical Mechanics: Theory and Experiment},
  volume={2021},
  number={7},
  pages={073101},
  year={2021},
  doi={10.1088/1742-5468/ac08fe},
  url={https://doi.org/10.1088/1742-5468/ac08fe}
}

@article{XCWu2021,
  title={Universal features of higher-form symmetries at phase transitions},
  author={Wu, Xiao-Chuan and Jian, Chao-Ming and Xu, Cenke},
  journal={SciPost Physics},
  volume={11},
  number={2},
  pages={033},
  year={2021},
  doi={10.21468/SciPostPhys.11.2.033},
  url={https://doi.org/10.21468/SciPostPhys.11.2.033}
}

@article{Harju1997StochasticGradient,
  title   = {Stochastic Gradient Approximation: An Efficient Method to Optimize Many-Body Wave Functions},
  author  = {Harju, A. and Barbiellini, B. and Siljam{\"a}ki, S. and Nieminen, R. M. and Ortiz, G.},
  journal = {Physical Review Letters},
  volume  = {79},
  number  = {6},
  pages   = {1173--1177},
  year    = {1997},
  doi     = {10.1103/PhysRevLett.79.1173}
}

@article{Sorella2005WaveFunctionOptimization,
  title   = {Wave Function Optimization in the Variational Monte Carlo Method},
  author  = {Sorella, Sandro},
  journal = {Physical Review B},
  volume  = {71},
  number  = {24},
  pages   = {241103},
  year    = {2005},
  doi     = {10.1103/PhysRevB.71.241103}
}

@article{Toulouse2007Optimization,
  title   = {Optimization of Quantum Monte Carlo Wave Functions by Energy Minimization},
  author  = {Toulouse, Julien and Umrigar, C. J.},
  journal = {The Journal of Chemical Physics},
  volume  = {126},
  number  = {8},
  pages   = {084102},
  year    = {2007},
  doi     = {10.1063/1.2437215}
}

@article{Carleo2017NeuralNetworkQuantum,
  title   = {Solving the Quantum Many-Body Problem with Artificial Neural Networks},
  author  = {Carleo, Giuseppe and Troyer, Matthias},
  journal = {Science},
  volume  = {355},
  number  = {6325},
  pages   = {602--606},
  year    = {2017},
  doi     = {10.1126/science.aag2302}
}

@article{luitz2015universal,
  title = {Universal logarithmic corrections to entanglement entropies in two dimensions with spontaneously broken continuous symmetries},
  author = {Luitz, David J. and Plat, Xavier and Alet, Fabien and Laflorencie, Nicolas},
  journal = {Phys. Rev. B},
  volume = {91},
  issue = {15},
  pages = {155145},
  numpages = {6},
  year = {2015},
  month = {Apr},
  publisher = {American Physical Society},
  doi = {10.1103/PhysRevB.91.155145},
  url = {https://link.aps.org/doi/10.1103/PhysRevB.91.155145}
}

@article{kulchytskyy2015detecting,
  title = {Detecting Goldstone modes with entanglement entropy},
  author = {Kulchytskyy, Bohdan and Herdman, C. M. and Inglis, Stephen and Melko, Roger G.},
  journal = {Phys. Rev. B},
  volume = {92},
  issue = {11},
  pages = {115146},
  numpages = {11},
  year = {2015},
  month = {Sep},
  publisher = {American Physical Society},
  doi = {10.1103/PhysRevB.92.115146},
  url = {https://link.aps.org/doi/10.1103/PhysRevB.92.115146}
}

@article{zhao2022scaling,
  title = {Scaling of Entanglement Entropy at Deconfined Quantum Criticality},
  author = {Zhao, Jiarui and Wang, Yan-Cheng and Yan, Zheng and Cheng, Meng and Meng, Zi Yang},
  journal = {Phys. Rev. Lett.},
  volume = {128},
  issue = {1},
  pages = {010601},
  numpages = {6},
  year = {2022},
  month = {Jan},
  publisher = {American Physical Society},
  doi = {10.1103/PhysRevLett.128.010601},
  url = {https://link.aps.org/doi/10.1103/PhysRevLett.128.010601}
}

@article{weber2021spin,
  title = {Spin versus bond correlations along dangling edges of quantum critical magnets},
  author = {Weber, Lukas and Wessel, Stefan},
  journal = {Phys. Rev. B},
  volume = {103},
  issue = {2},
  pages = {L020406},
  numpages = {5},
  year = {2021},
  month = {Jan},
  publisher = {American Physical Society},
  doi = {10.1103/PhysRevB.103.L020406},
  url = {https://link.aps.org/doi/10.1103/PhysRevB.103.L020406}
}

@article{wang2026boundary,
  title = {Boundary renormalization group flow of entanglement entropy at a ($2+1$)-dimensional quantum critical point},
  author = {Wang, Zhiyan and Wang, Zhe and Ding, Yi-Ming and Liu, Zenan and Yan, Zheng and Zhang, Long},
  journal = {Phys. Rev. B},
  volume = {113},
  issue = {16},
  pages = {L161104},
  numpages = {6},
  year = {2026},
  month = {Apr},
  publisher = {American Physical Society},
  doi = {10.1103/yv52-wm2s},
  url = {https://link.aps.org/doi/10.1103/yv52-wm2s}
}

@article{metlitski2009entanglement,
  title = {Entanglement entropy in the $O(N)$ model},
  author = {Metlitski, Max A. and Fuertes, Carlos A. and Sachdev, Subir},
  journal = {Phys. Rev. B},
  volume = {80},
  issue = {11},
  pages = {115122},
  numpages = {20},
  year = {2009},
  month = {Sep},
  publisher = {American Physical Society},
  doi = {10.1103/PhysRevB.80.115122},
  url = {https://link.aps.org/doi/10.1103/PhysRevB.80.115122}
}

@article{whitsitt2017entanglement,
  title = {Entanglement entropy of large-$N$ Wilson-Fisher conformal field theory},
  author = {Whitsitt, Seth and Witczak-Krempa, William and Sachdev, Subir},
  journal = {Phys. Rev. B},
  volume = {95},
  issue = {4},
  pages = {045148},
  numpages = {9},
  year = {2017},
  month = {Jan},
  publisher = {American Physical Society},
  doi = {10.1103/PhysRevB.95.045148},
  url = {https://link.aps.org/doi/10.1103/PhysRevB.95.045148}
}

@article{calabrese2004entanglement,
   title={Entanglement entropy and quantum field theory},
   volume={2004},
   ISSN={1742-5468},
   url={http://dx.doi.org/10.1088/1742-5468/2004/06/P06002},
   DOI={10.1088/1742-5468/2004/06/p06002},
   number={06},
   journal={Journal of Statistical Mechanics: Theory and Experiment},
   publisher={IOP Publishing},
   author={Pasquale Calabrese and John Cardy},
   year={2004},
   month=June, pages={P06002} }

@article{holzhey1994geometric,
   title={Geometric and renormalized entropy in conformal field theory},
   volume={424},
   ISSN={0550-3213},
   url={http://dx.doi.org/10.1016/0550-3213(94)90402-2},
   DOI={10.1016/0550-3213(94)90402-2},
   number={3},
   journal={Nuclear Physics B},
   publisher={Elsevier BV},
   author={Holzhey, Christoph and Larsen, Finn and Wilczek, Frank},
   year={1994},
   month=Aug, pages={443–467} }

@article{kallin2011anomalies,
  title = {Anomalies in the entanglement properties of the square-lattice Heisenberg model},
  author = {Kallin, Ann B. and Hastings, Matthew B. and Melko, Roger G. and Singh, Rajiv R. P.},
  journal = {Phys. Rev. B},
  volume = {84},
  issue = {16},
  pages = {165134},
  numpages = {11},
  year = {2011},
  month = {Oct},
  publisher = {American Physical Society},
  doi = {10.1103/PhysRevB.84.165134},
  url = {https://link.aps.org/doi/10.1103/PhysRevB.84.165134}
}

@article{song2024extracting,
   title={Extracting subleading corrections in entanglement entropy at quantum phase transitions},
   volume={17},
   ISSN={2542-4653},
   url={http://dx.doi.org/10.21468/SciPostPhys.17.1.010},
   DOI={10.21468/scipostphys.17.1.010},
   number={1},
   journal={SciPost Physics},
   publisher={Stichting SciPost},
   pages = {010},
   author={Song, Menghan and Zhao, Jiarui and Meng, Zi Yang and Xu, Cenke and Cheng, Meng},
   year={2024},
   month=July }

@article{matsumoto2002ground,
  title = {Ground-state phase diagram of quantum Heisenberg antiferromagnets on the anisotropic dimerized square lattice},
  author = {Matsumoto, Munehisa and Yasuda, Chitoshi and Todo, Synge and Takayama, Hajime},
  journal = {Phys. Rev. B},
  volume = {65},
  issue = {1},
  pages = {014407},
  numpages = {8},
  year = {2001},
  month = {Nov},
  publisher = {American Physical Society},
  doi = {10.1103/PhysRevB.65.014407},
  url = {https://link.aps.org/doi/10.1103/PhysRevB.65.014407}
}

@article{weber2018nonordinary,
  title = {Nonordinary edge criticality of two-dimensional quantum critical magnets},
  author = {Weber, Lukas and Parisen Toldin, Francesco and Wessel, Stefan},
  journal = {Phys. Rev. B},
  volume = {98},
  issue = {14},
  pages = {140403(R)},
  numpages = {5},
  year = {2018},
  month = {Oct},
  publisher = {American Physical Society},
  doi = {10.1103/PhysRevB.98.140403},
  url = {https://link.aps.org/doi/10.1103/PhysRevB.98.140403}
}

@article{Grover2011Entanglement,
  title = {Entanglement of Interacting Fermions in Quantum Monte Carlo Calculations},
  author = {Grover, Tarun},
  journal = {Phys. Rev. Lett.},
  volume = {111},
  issue = {13},
  pages = {130402},
  numpages = {5},
  year = {2013},
  month = {Sep},
  publisher = {American Physical Society},
  doi = {10.1103/PhysRevLett.111.130402},
  url = {https://link.aps.org/doi/10.1103/PhysRevLett.111.130402}
}

@article{Hastings2010Measuring,
  title = {Measuring Renyi Entanglement Entropy in Quantum Monte Carlo Simulations},
  author = {Hastings, Matthew B. and Gonz{\'a}lez, Iv{\'a}n and Kallin, Ann B. and Melko, Roger G.},
  journal = {Phys. Rev. Lett.},
  volume = {104},
  issue = {15},
  pages = {157201},
  numpages = {4},
  year = {2010},
  month = {Apr},
  publisher = {American Physical Society},
  doi = {10.1103/PhysRevLett.104.157201},
  url = {https://link.aps.org/doi/10.1103/PhysRevLett.104.157201}
}

\onecolumngrid
\begin{center}
    \textbf{End Matter}
\end{center}
\twocolumngrid

\setcounter{equation}{0}
\renewcommand{\theequation}{A\arabic{equation}}
\renewcommand{\theHequation}{EM.A.\arabic{equation}}





\emph{Learning the lattice Bisognano--Wichmann scale.---}
We apply the learning framework to determine the model-dependent overall prefactor of the lattice Bisognano--Wichmann (LBW) entanglement Hamiltonian. 
This prefactor is generally not known a priori.

The LBW ansatz prescribes the spatial dependence of the EH couplings, while their overall scale is set by the prefactor $2\pi/v$, where $v$ is the low-energy velocity.
Conventionally, $v$ is supplied as an external input from analytical or independent numerical calculations in some solvable cases~\cite{sen2015velocity}.
Ref.~\cite{yang2026limits} instead extracted this scale by matching the imaginary-time correlation functions of the LBW EHs and exact EHs, but the precision is hard to control.
Here we determine this scale by matching nearest-neighbor correlations while keeping the prescribed spatial profile fixed. 
The learned scale is expressed as $2\pi/v_{\rm eff}(L)$, where $v_{\rm eff}(L)$ denotes the effective velocity for a given system size and geometry.

We consider equal bipartitions of $1$D Heisenberg chains with $2L$ sites and $2$D Heisenberg systems of size $2L\times L$, giving subsystems of length $L$ and size $L\times L$, respectively. 
The spatial profiles depend on the geometry and boundary conditions of the original system. 
We denote the horizontal and vertical coupling profiles by $\Gamma_x$ and $\Gamma_y$, respectively, with the common prefactor factored out and the original exchange coupling set to unity.

For the $1$D chains, we use the finite-system conformal extensions of the BW profile~\cite{cardy2016entanglement,giudici2018entanglement}.
For PBC, the coupling profile is
\begin{equation}
    \Gamma_x = \frac{L}{\pi}\sin\left(\frac{\pi x}{L}\right).
\end{equation}
For OBC, it is
\begin{equation}
    \Gamma_x = \frac{2L}{\pi}\sin\left(\frac{\pi x}{2L}\right).
\end{equation}
Here $x=1,\ldots,L-1$ labels the bonds within subsystem $A$, measured from the entanglement boundary.

For the $2$D cylinder, the original system has OBC along $x$ and PBC along $y$.
We use the linear BW profile~\cite{giudici2018entanglement}, 
\begin{equation}
    \Gamma_x = x,
    \qquad
    \Gamma_y = x-\frac12.
\end{equation}
For the torus, the original system has PBC along both directions, and we use the parabolic ansatz~\cite{giudici2018entanglement}, 
\begin{equation}
    \Gamma_x = \frac{x(L-x)}{L},
    \qquad
    \Gamma_y = \frac{\left(x-\frac12\right)\left(L-x+\frac12\right)}{L}.
\end{equation}
Here $x=1,\ldots,L-1$ for horizontal bonds and $x=1,\ldots,L$ for vertical bonds, with bond centers at distances $x$ and $x-\frac12$ from the entanglement cut, respectively.

The target observables are the nearest-neighbor correlations of the original system restricted to subsystem $A$. 
In $1$D, these correlations are obtained using DMRG~\cite{white1992density}, while in $2$D they are measured using SSE QMC at sufficiently low temperatures to approximate the ground state. 
The target correlations remain fixed throughout optimization, and the trial LBW Hamiltonian is simulated on subsystem $A$ at unit effective inverse temperature.

At each learning iteration, the trial Hamiltonian is equilibrated and sampled with the current prefactor. 
The measured correlations and their responses determine the loss gradient through the SSE estimators derived above. 
Since the prefactor rescales all couplings simultaneously, the response of each correlation is evaluated from its covariance with the total SSE operator count. 
The contributions from all correlation mismatches are combined to update the common prefactor by gradient descent, preserving the prescribed spatial profile throughout optimization.

\begin{table}[t]
    \caption{
    Effective velocities $v_{\rm eff}(L)$ obtained by learning the LBW prefactor from equal-time nearest-neighbor correlations.
    The subsystem contains $L$ sites in $1$D and $L\times L$ sites in $2$D.
    Bulk reference velocities are included for comparison.
    }
    \label{tab:velocity}
    \centering
    \setlength{\tabcolsep}{8pt}
    \begin{tabular}{lcccc}
    \hline\hline
    $L$ & $1$D OBC & $1$D PBC & $2$D cylinder & $2$D torus \\
    \hline
    4  & 1.513(4) & 1.634(3) & 1.685(4) & 1.454(4) \\
    12 & 1.544(6) & 1.578(7) & 1.701(3) & 1.623(2) \\
    20 & 1.552(8) & 1.568(7) & 1.705(4) & 1.656(2) \\
    \hline
    $v_{\rm bulk}$ & $\pi/2$ & $\pi/2$ & 1.657(2) & 1.657(2) \\
    \hline\hline
    \end{tabular}
\end{table}

Representative results are summarized in Table~\ref{tab:velocity}.
For the spin-$1/2$ Heisenberg chain, the learned effective velocities are close to the exact bulk value $v=\pi/2$~\cite{des1962spin}.
In particular, the PBC result at $L=20$ agrees with this value within the quoted uncertainty.
In $2$D, the torus result at $L=20$, $v_{\rm eff}=1.656(2)$, agrees with the numerical bulk reference $v=1.657(2)$~\cite{beard1998square}.
The cylinder results exhibit a modest size dependence and remain approximately $3\%$ above the bulk reference at larger sizes.

The finite-size and geometry-dependent deviations indicate that the learned effective velocity can also absorb discrepancies between the prescribed LBW profile and the exact lattice EH. 
These results establish a route for determining the LBW scale directly from equal-time correlations.
The scale is adapted to each finite system, while the bulk velocities provide independent references for assessing its size and geometry dependence.

\clearpage
\onecolumngrid
\newpage
\setcounter{page}{1}
\setcounter{equation}{0}
\setcounter{figure}{0}
\setcounter{table}{0}
\renewcommand{\theequation}{S\arabic{equation}}
\renewcommand{\thefigure}{S\arabic{figure}}
\renewcommand{\thetable}{S\Roman{table}}
\renewcommand{\theHequation}{SM.\arabic{equation}}
\renewcommand{\theHfigure}{SM.\arabic{figure}}
\renewcommand{\theHtable}{SM.\arabic{table}}
\hypersetup{pageanchor=false}

\begin{center}
    \textbf{Supplemental Material for}\\
    \textbf{Observable-Based Learning of Effective Entanglement Hamiltonians}
    \\[1em]
    Siyi Yang, Yuzhi Liu, Yi-Ming Ding, and Zheng Yan
\end{center}

\section{Observable response and learning gradient}
\label{sec:sm_gradient}

We derive the observable response and loss gradient underlying the learning procedure.
For the parametrized trial entanglement Hamiltonian (EH)
\begin{equation}
    \tilde H_A(\bm J)=\sum_i J_i h_i,
    \label{eq:sm_trial_h}
\end{equation}
the normalized density matrix is
\begin{equation}
    \tilde\rho_A(\bm J)=\frac{e^{-\tilde H_A(\bm J)}}{\tilde Z_A(\bm J)},\qquad \tilde Z_A(\bm J)=\mathrm{Tr}_A e^{-\tilde H_A(\bm J)}.
    \label{eq:sm_rho}
\end{equation}
The effective inverse temperature is absorbed into $\tilde H_A$, so the imaginary-time interval has unit length.

For an observable $O$ with no explicit dependence on $J_i$,
\begin{equation}
    \langle O\rangle_{\bm J}=\frac{\mathrm{Tr}_A[e^{-\tilde H_A}O]}{\tilde Z_A}.
    \label{eq:sm_expectation}
\end{equation}
Since $\tilde H_A$ and $\partial\tilde H_A/\partial J_i=h_i$ do not generally commute, we use the Duhamel identity
\begin{equation}
    \frac{\partial e^{-\tilde H_A}}{\partial J_i}=-\int_0^1 d\tau\,e^{-(1-\tau)\tilde H_A}h_i e^{-\tau\tilde H_A}.
    \label{eq:sm_duhamel}
\end{equation}

Applying Eq.~\eqref{eq:sm_duhamel} to the partition function and using cyclicity of the trace gives
\begin{equation}
    \frac{\partial\tilde Z_A}{\partial J_i}=-\int_0^1 d\tau\,\mathrm{Tr}_A[e^{-(1-\tau)\tilde H_A}h_i e^{-\tau\tilde H_A}]=-\tilde Z_A\langle h_i\rangle_{\bm J}.
    \label{eq:sm_dZ}
\end{equation}
For the numerator,
\begin{equation}
    \frac{\partial}{\partial J_i}\mathrm{Tr}_A[e^{-\tilde H_A}O]=-\tilde Z_A\int_0^1 d\tau\,\langle h_i(\tau)O\rangle_{\bm J},
    \label{eq:sm_dN}
\end{equation}
where $h_i(\tau)=e^{\tau\tilde H_A}h_i e^{-\tau\tilde H_A}$.
Using the quotient rule then yields
\begin{equation}
    \frac{\partial\langle O\rangle_{\bm J}}{\partial J_i}=-\int_0^1 d\tau\left[\langle h_i(\tau)O\rangle_{\bm J}-\langle h_i\rangle_{\bm J}\langle O\rangle_{\bm J}\right].
    \label{eq:sm_response}
\end{equation}
This reproduces Eq.~\eqref{eq:kubo_response} of the main text.

Equation~\eqref{eq:sm_response} generalizes the conventional covariance response to noncommuting quantum operators.
If $[h_i,\tilde H_A]=0$, then $h_i(\tau)=h_i$, and the imaginary-time integral becomes trivial,
\begin{equation}
    \frac{\partial\langle O\rangle_{\bm J}}{\partial J_i}=-\left[\langle h_iO\rangle_{\bm J}-\langle h_i\rangle_{\bm J}\langle O\rangle_{\bm J}\right].
    \label{eq:sm_commuting_response}
\end{equation}
Thus, the ordinary covariance form is recovered when $h_i$ commutes with $\tilde H_A$, whereas the full imaginary-time integral is generally required otherwise.

Defining the observable mismatch 
$\Delta O_\mu=\langle O_\mu\rangle_{\bm J}-\langle O_\mu\rangle_{\rm tar}$, 
the quadratic loss is
$\mathcal L=\frac12\sum_\mu w_\mu(\Delta O_\mu)^2$.
Its gradient follows from the chain rule,
\begin{equation}
    \frac{\partial\mathcal L}{\partial J_i}=\sum_\mu w_\mu\Delta O_\mu\frac{\partial\langle O_\mu\rangle_{\bm J}}{\partial J_i}.
    \label{eq:sm_loss_gradient}
\end{equation}
The learning gradient is therefore determined by the observable mismatch and the response matrix of the trial ensemble.

\section{SSE evaluation of the response matrix}
\label{sec:sm_sse}

We evaluate the trial ensemble using the stochastic series expansion (SSE). 
Here we show how the imaginary-time fluctuation relation derived above is evaluated through operator-count correlations.

For the trial EH $\tilde H_A=\sum_i J_i h_i$, the partition function admits the Taylor expansion
\begin{equation}
    \tilde Z_A
    =
    \sum_\alpha
    \sum_{n=0}^{\infty}
    \frac{1}{n!}
    \langle\alpha|(-\tilde H_A)^n|\alpha\rangle.
    \label{eq:sm_sse_partition}
\end{equation}
In the SSE representation, the Hamiltonian is decomposed into diagonal and off-diagonal bond operators in the computational basis~\cite{sandvik1999stochastic,sandvik2010computational}.
The expansion is truncated at a sufficiently large order $M$ and represented by an operator string of fixed length $M$, padded with identity operators.
The expansion order $n$ then counts the number of nonidentity operators in the string.

Each Hamiltonian operator in the string carries the coupling constant of its corresponding term.
Let $n_i$ denote the number of nonidentity operators associated with the coupling $J_i$.
If one variational parameter controls several bonds, $n_i$ counts all corresponding bond operators.
For a sampled SSE configuration $C$, the dependence of its weight on the couplings takes the form
\begin{equation}
    W(C;\bm J)
    \propto
    \prod_i J_i^{n_i(C)},
    \label{eq:sm_sse_weight}
\end{equation}
where the remaining factors are independent of $\bm J$.
Consequently,
\begin{equation}
    \frac{\partial W(C;\bm J)}{\partial J_i}
    =
    \frac{n_i(C)}{J_i}W(C;\bm J).
    \label{eq:sm_sse_weight_derivative}
\end{equation}

For an observable $O$ with a coupling-independent SSE estimator $O(C)$, its expectation value is
\begin{equation}
    \langle O\rangle_{\bm J}
    =
    \frac{
        \sum_C W(C;\bm J)O(C)
    }{
        \sum_C W(C;\bm J)
    }.
    \label{eq:sm_sse_observable}
\end{equation}
Differentiating with respect to $J_i$ and using Eq.~\eqref{eq:sm_sse_weight_derivative}, we obtain
\begin{equation}
    \begin{aligned}
    \frac{\partial\langle O\rangle_{\bm J}}{\partial J_i}
    &=
    \frac{1}{J_i}
    \left[
        \langle n_iO\rangle_{\bm J}
        -
        \langle n_i\rangle_{\bm J}
        \langle O\rangle_{\bm J}
    \right].
    \end{aligned}
    \label{eq:sm_sse_covariance}
\end{equation}
The response is therefore determined by the covariance between the operator count $n_i$ and the observable $O$, both measured in the same SSE ensemble.

On the other hand, the imaginary-time fluctuation relation derived in Eq.~\eqref{eq:sm_response} gives
\begin{equation}
    \frac{\partial\langle O\rangle_{\bm J}}{\partial J_i}
    =
    -\int_0^1 d\tau\,
    \left[
        \langle h_i(\tau)O\rangle_{\bm J}
        -
        \langle h_i\rangle_{\bm J}
        \langle O\rangle_{\bm J}
    \right].
    \label{eq:sm_sse_kubo}
\end{equation}
Comparing the two expressions, we obtain
\begin{equation}
    \int_0^1 d\tau\,
    \left[
        \langle h_i(\tau)O\rangle_{\bm J}
        -
        \langle h_i\rangle_{\bm J}
        \langle O\rangle_{\bm J}
    \right]
    =
    -\frac{
        \langle n_iO\rangle_{\bm J}
        -
        \langle n_i\rangle_{\bm J}
        \langle O\rangle_{\bm J}
    }{J_i}.
    \label{eq:sm_sse_connected}
\end{equation}
Thus, the imaginary-time integrated response can be evaluated directly from operator-count correlations, without explicitly sampling the imaginary-time dependence of the correlation function.

In practical SSE implementations, constant shifts may be introduced in the diagonal bond operators to obtain nonnegative matrix elements.
These shifts contribute only an overall factor to the partition function and cancel in normalized expectation values, leaving the covariance response unchanged.

For the learning observables $\{O_\mu\}$, the response matrix is therefore
\begin{equation}
    \frac{\partial\langle O_\mu\rangle_{\bm J}}{\partial J_i}
    =
    \frac{
        \langle n_iO_\mu\rangle_{\bm J}
        -
        \langle n_i\rangle_{\bm J}
        \langle O_\mu\rangle_{\bm J}
    }{J_i}.
    \label{eq:sm_response_matrix}
\end{equation}

For the logarithmic parametrization
$\theta_i=\ln J_i$, the factor $J_i$ cancels,
\begin{equation}
    \frac{\partial\langle O_\mu\rangle_{\bm J}}
    {\partial\theta_i}
    =
    \langle n_iO_\mu\rangle_{\bm J}
    -
    \langle n_i\rangle_{\bm J}
    \langle O_\mu\rangle_{\bm J}.
    \label{eq:sm_sse_log_response}
\end{equation}
Thus, the response with respect to the logarithmic coupling is given directly by the covariance between the operator count and the learning observable.

\section{Optimization and statistical sampling}

At each learning iteration, the variational couplings are held fixed while the SSE Markov chain is equilibrated and sampled.
The target observables are obtained independently from the original many-body system and remain fixed throughout the optimization.
The sampled observables and operator-count correlations determine the loss gradient through Eqs.~\eqref{eq:sm_loss_gradient} and \eqref{eq:sm_response_matrix}.
The couplings are then updated, and the procedure is repeated until convergence.

For simple optimization problems, gradient descent can be applied directly to the couplings $J_i$,
\begin{equation}
    J_i^{(t+1)}
    =
    J_i^{(t)}
    -
    \eta
    \frac{\partial\mathcal L}{\partial J_i},
    \label{eq:sm_gd_update}
\end{equation}
where $\eta$ is the learning rate.

When the couplings span substantially different magnitudes, we use the logarithmic parametrization
\begin{equation}
    \theta_i=\ln J_i,\qquad J_i=e^{\theta_i}.
    \label{eq:sm_log_parameter}
\end{equation}
An additive change in $\theta_i$ corresponds to a multiplicative update in $J_i$, providing a more uniform optimization scale while preserving $J_i>0$.
Combining Eqs.~\eqref{eq:sm_loss_gradient} and \eqref{eq:sm_sse_log_response}, the loss gradient with respect to $\theta_i$ is
\begin{equation}
    g_i
    =
    \frac{\partial\mathcal L}{\partial\theta_i}
    =
    \sum_\mu w_\mu\Delta O_\mu
    \left[
        \langle n_iO_\mu\rangle_{\bm J}
        -
        \langle n_i\rangle_{\bm J}
        \langle O_\mu\rangle_{\bm J}
    \right].
    \label{eq:sm_log_loss_gradient}
\end{equation}

For optimizations involving multiple parameters and stochastic gradients, we employ the Adam optimizer~\cite{kingma2014adam}, illustrated here for the logarithmic parametrization.
Adam adaptively rescales parameter updates using exponential moving averages of the stochastic gradients and their squares.
At iteration $t$, these moments are updated as
\begin{equation}
    \begin{aligned}
    m_i^{(t)}
    &=
    \beta_1m_i^{(t-1)}
    +(1-\beta_1)g_i^{(t)},\\
    v_i^{(t)}
    &=
    \beta_2v_i^{(t-1)}
    +(1-\beta_2)[g_i^{(t)}]^2,
    \end{aligned}
    \label{eq:sm_adam_moments}
\end{equation}
with $m_i^{(0)}=v_i^{(0)}=0$.
The bias-corrected moments are
\begin{equation}
    \hat m_i^{(t)}
    =
    \frac{m_i^{(t)}}{1-\beta_1^t},
    \qquad
    \hat v_i^{(t)}
    =
    \frac{v_i^{(t)}}{1-\beta_2^t},
    \label{eq:sm_adam_bias}
\end{equation}
and the parameters are updated according to
\begin{equation}
    \theta_i^{(t+1)}
    =
    \theta_i^{(t)}
    -
    \eta
    \frac{\hat m_i^{(t)}}
    {\sqrt{\hat v_i^{(t)}}+\epsilon}.
    \label{eq:sm_adam_update}
\end{equation}
The couplings are then recovered as $J_i^{(t+1)}=e^{\theta_i^{(t+1)}}$.

Because the gradients are estimated from finite QMC samples, the learning trajectories are stochastic~\cite{bottou2018optimization}.
We perform independent learning realizations, each with its own QMC sampling and parameter optimization.
The resulting learned couplings are averaged across realizations, and the variation among realizations is used to estimate their statistical uncertainties.
The choice of parametrization and optimizer, together with the sampling and averaging procedures, is specified for each calculation in the following sections.

\section{Benchmarks on Hamiltonians with known couplings}

We first validate the learning procedure by recovering known coupling parameters from equal-time correlations. 
For each benchmark, the target correlations are obtained from an independent SSE simulation of a Hamiltonian with prescribed couplings and held fixed throughout optimization. 
The trial Hamiltonian is then sampled at each learning iteration, and its couplings are updated by gradient descent using the SSE response estimators derived above.

\begin{figure}[t]
    \centering
    \includegraphics[width=\textwidth]{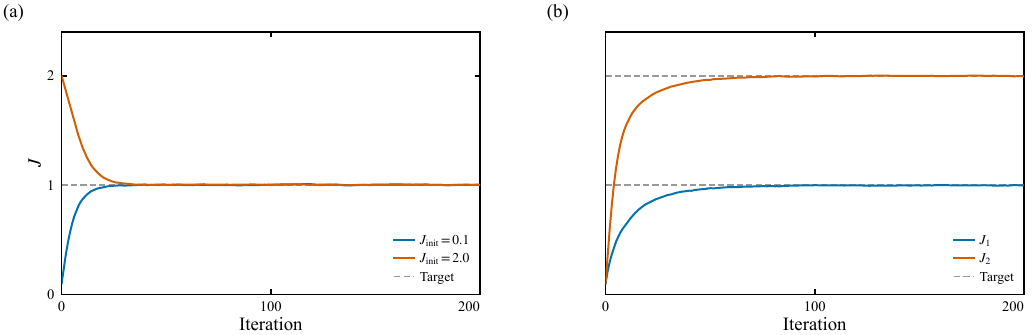}
    \caption{
    Learning trajectories for Hamiltonians with known couplings at inverse temperature $\beta=1$.
    (a) Spin-$1/2$ Heisenberg chain with $L=4$ and PBC, starting from $J_{\rm init}=0.1$ and $2.0$.
    (b) $4\times4$ dimerized Heisenberg model on a cylinder with OBC along $x$ and PBC along $y$, where $J_1$ and $J_2$ are optimized simultaneously from $J_1=J_2=0.1$.
    Dashed lines indicate the target coupling values.
    Only the first 200 optimization iterations are shown.
    }
    \label{fig:benchmarks}
\end{figure}

For the one-parameter benchmark, we consider a spin-$1/2$ Heisenberg chain with $L=4$ and periodic boundary conditions (PBC) at inverse temperature $\beta=1$.
The learning observable is the nearest-neighbor spin correlation, averaged over all bonds. 
Starting from $J_{\mathrm{init}}=0.1$ and $2.0$, on opposite sides of the target coupling $J_{\mathrm{target}}=1$, both learning trajectories converge to the prescribed value, as shown in Fig.~\ref{fig:benchmarks}(a).

For the two-parameter benchmark, we consider a $4\times4$ dimerized Heisenberg model on a cylinder, with open boundary conditions (OBC) along the $x$ direction.
The two bond couplings are optimized simultaneously using the full response matrix of the two corresponding nearest-neighbor correlations.
Starting from $J_1=J_2=0.1$, both couplings converge to their target values $J_1=1$ and $J_2=2$, as shown in Fig.~\ref{fig:benchmarks}(b).

For both benchmarks, the small number of parameters and their comparable magnitudes allow us to update the physical couplings directly by gradient descent, without the logarithmic parametrization or Adam optimizer used in subsequent free-form learning calculations.

These benchmarks demonstrate that the SSE response estimators and gradient-based updates recover prescribed Hamiltonian parameters in both single- and multi-parameter optimization problems, providing a numerical validation of the learning procedure before its application to effective EHs.

\section{Free-form learning and entropy evaluation}

\subsection{Free-form EH parametrization and optimization}

The free-form EH is parametrized using nearest-neighbor Heisenberg interactions as the operator basis and the corresponding equal-time correlations as the learning observables,
\begin{equation}
    \tilde H_A(\bm J)
    =
    \sum_{\langle ij\rangle\subset A}
    J_{ij}\bm S_i\cdot\bm S_j .
\end{equation}
The target correlations are obtained from DMRG in $1$D and from sufficiently low-temperature SSE simulations of the original $2$D systems.
Only bonds entirely contained within subsystem $A$ are included.
These target observables are prepared before learning and remain fixed throughout the optimization.
The trial EH is sampled on subsystem $A$ at unit effective inverse temperature.

The coupling parametrization incorporates selected spatial symmetries without assuming a functional form for the coupling profile.
In $1$D, the $L-1$ nearest-neighbor couplings of a subsystem of length $L$ are optimized independently.
For the square-lattice systems, translation symmetry along the periodic $y$ direction is used to group equivalent bonds.
Horizontal and vertical bonds are treated as independent coupling groups.
For the torus geometry, additional reflection symmetry further identifies equivalent coupling groups.
These symmetry constraints reduce the number of independent parameters without prescribing their spatial dependence.

Let $\mathcal B_a$ denote the set of bonds sharing the same variational coupling $J_a$, such that $J_{ij}=J_a$ for all $\langle ij\rangle\in\mathcal B_a$.
For each coupling group $a$, we define the group-averaged correlation in the trial ensemble as
\begin{equation}
    C_a
    =
    \frac{1}{|\mathcal B_a|}
    \sum_{\langle ij\rangle\in\mathcal B_a}
    \langle\bm S_i\cdot\bm S_j\rangle_{\bm J},
\end{equation}
with the corresponding target value denoted by $C_a^{\rm tar}$.

The independent couplings are optimized using the logarithmic parametrization $\theta_a=\ln J_a$ to preserve positive exchange values.
In the SSE representation, the diagonal-response update direction is
\begin{equation}
    g_a^{\rm diag}
    =
    (C_a-C_a^{\rm tar})
    \left[
    \langle n_a\widehat C_a\rangle_{\bm J}
    -
    \langle n_a\rangle_{\bm J}
    \langle\widehat C_a\rangle_{\bm J}
    \right],
\end{equation}
where $\widehat C_a$ is the coupling-independent SSE estimator satisfying $C_a=\langle\widehat C_a\rangle_{\bm J}$, and $n_a$ counts all nonidentity operators associated with the coupling group.
In practice, we use the diagonal spin correlations $\langle S_i^z S_j^z\rangle$ for observable matching. 
Owing to $SU(2)$ symmetry, $\langle\bm S_i\cdot\bm S_j\rangle=3\langle S_i^zS_j^z\rangle$, so matching either set of correlations yields the same optimized couplings.
The observable weights are set to unity for all groups.
This update retains only the response of each group-averaged correlation to its associated coupling, providing a diagonal-response approximation to the full loss gradient in Eq.~\eqref{eq:sm_log_loss_gradient}.
This approximation reduces the number of response elements required in the many-parameter optimization.
The resulting update directions are used in Adam to optimize the logarithmic parameters, with bounded updates to avoid excessively large changes in the physical couplings.
The SSE configuration is retained between iterations and re-equilibrated after each coupling update.

\subsection{Sequential-freezing strategy}

Spatially varying EH couplings can converge at different rates, so continuing to update stabilized parameters may introduce stochastic drift.
For the uniform Heisenberg systems, we find that the couplings tend to stabilize from the entanglement boundary toward the subsystem interior.
We therefore employ a sequential-freezing strategy, starting with all couplings active and progressively freezing stabilized couplings from the boundary inward.
A coupling is frozen when the relative variation of its physical coupling $J_a$ over a chosen number of optimization iterations falls below a specified tolerance.
The convergence window and tolerance are adjusted according to the optimization problem.
Frozen couplings remain part of the simulated EH but are excluded from subsequent updates.
For bipartitions with two equivalent entanglement boundaries, freezing proceeds symmetrically from both boundaries toward the center.

Although the trial correlations may approach their target values before the coupling parameters fully converge, optimization continues until the parameters become stable.
The final learned correlations are then checked against their target values to verify the quality of the optimization.

\begin{figure*}[t]
    \centering
    \includegraphics[width=\textwidth]{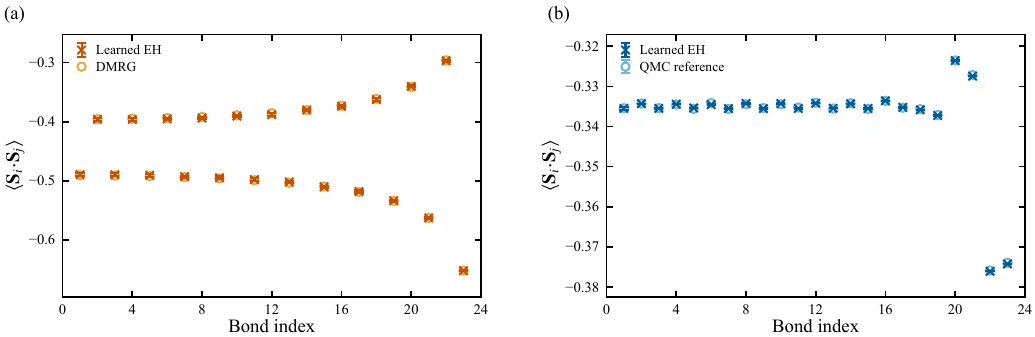}
    \caption{
    Comparison between target nearest-neighbor correlations and those obtained from the learned EHs.
    (a) Representative $1$D result for an OBC Heisenberg chain with subsystem size $L=24$, where the target correlations are obtained from DMRG.
    (b) Representative $2$D result for a cylinder geometry with subsystem size $L\times L=12\times12$, where the target correlations are obtained from low-temperature SSE simulations.
    The correlations are shown for the bond groups used in the free-form parametrization.
    The learned EH correlations are evaluated after fixing the optimized couplings.
    Error bars denote Monte Carlo statistical uncertainties where applicable.
    }
    \label{fig:freeform_corr_match}
\end{figure*}

Fig.~\ref{fig:freeform_corr_match} compares the target nearest-neighbor correlations with those obtained from the learned EHs for representative $1$D and $2$D systems.
The learned EHs reproduce the target correlations within small residual deviations.

\subsection{Entropy evaluation}

After learning, the couplings are averaged over independent learning realizations to construct a fixed EH for each system size.
Its von Neumann and second R\'enyi entropies are evaluated from thermodynamic quantities at unit effective inverse temperature through Eq.~\eqref{eq:freeform_entropies} of the main text.
The required partition functions are obtained using thermal reweight-annealing in SSE~\cite{ding2024reweight}, following the procedure previously applied to LBW entanglement Hamiltonians.
The learned couplings remain fixed while the effective inverse temperature $\beta_A$ is varied.

In the SSE implementation, the Heisenberg Hamiltonian includes the usual constant shift $-\frac14\sum_{\langle ij\rangle\subset A}J_{ij}$.
Throughout this subsection, we use $\tilde H_A$ and $\tilde Z_A$ in this shifted convention.
The shift leaves the normalized density matrix and the entropies unchanged when the partition functions and mean energy are evaluated consistently.
For two nearby effective inverse temperatures, the partition function ratio is
\begin{equation}
    \frac{\tilde Z_A(\beta_A')}{\tilde Z_A(\beta_A)}
    =
    \left\langle
        \left(\frac{\beta_A'}{\beta_A}\right)^m
    \right\rangle_{\beta_A},
    \label{eq:sm_entropy_reweight}
\end{equation}
where $m$ denotes the total number of nonidentity SSE operators, including both diagonal and off-diagonal bond operators.
The total expansion order enters because varying $\beta_A$ rescales the entire fixed EH.

The reweighting is performed along two inverse-temperature paths, from $\beta_A=1$ toward the infinite-temperature limit and from $\beta_A=2$ to $1$.
Following the pseudo-automatic annealing scheme of Ref.~\cite{ding2024reweight}, successive inverse temperatures are chosen according to $\beta_{k-1}=\epsilon^{1/(\Lambda\beta_k N_A)}\beta_k$, where $N_A$ denotes the number of spins in subsystem $A$, $\Lambda$ is a prescribed parameter used to estimate the mean SSE expansion order, and $\epsilon<1$ controls the nominal reweighting ratio.
Near the infinite-temperature limit, a lower bound is imposed on the ratio $\beta_{k-1}/\beta_k$ to avoid numerical underflow. 
The first path is connected to the exactly known infinite-temperature reference $\tilde Z_A(0)=2^{N_A}$ through the final reweighting step. 
Accumulating the logarithms of the measured partition-function ratios then gives $\ln\tilde Z_A(1)$.
The second path determines $\ln[\tilde Z_A(1)/\tilde Z_A(2)]$ and hence $\ln\tilde Z_A(2)$.

The mean energy required for the von Neumann entropy is obtained from a separate, extended SSE measurement at $\beta_A=1$, using
\begin{equation}
    \bigl\langle\tilde H_A(\bm J)\bigr\rangle_{\beta_A=1}
    =
    -\langle m\rangle_{\beta_A=1}.
    \label{eq:sm_entropy_energy}
\end{equation}
The partition functions and mean energy then determine $S_{\rm vN}$ and $S_2$ through Eq.~\eqref{eq:freeform_entropies} of the main text.

The reference entropies are obtained independently from the original systems.
In $1$D, both $S_{\rm vN}$ and $S_2$ are calculated using DMRG~\cite{white1992density}.
In $2$D, the QMC reference for $S_2$ is obtained using the nonequilibrium replica QMC method~\cite{d2020entanglement,zhao2022measuring}.
A parameter $\lambda$ interpolates between two disconnected replicas and replicas cyclically joined in imaginary time on subsystem $A$.
The corresponding partition-function ratio is evaluated by accumulating reweighting factors along the interpolation.
For the $2$D QMC reference calculations, we set the inverse temperature of the original system to $\beta=4L$ to approximate the ground state.
The $2$D $S_{\rm vN}$ results are obtained from the learned EHs, without independent reference data.

Independent Monte Carlo realizations are used to estimate the statistical uncertainties of the entropy measurements for each fixed learned EH.
These uncertainties reflect sampling fluctuations and do not include variations between independent learning runs.
Since neither entropy enters the learning objective, comparison with the independent reference results tests whether matching local correlations also captures the entanglement properties of the target state.

\section{EH learning at the dimerized quantum critical point}

\subsection{Lattice geometry and EH parametrization}

We consider the columnar dimerized spin-$1/2$ Heisenberg model on a $2L\times L$ cylinder, with open boundaries along $x$ and periodic boundaries along $y$.
Strong and weak bonds alternate along the horizontal direction, while all vertical bonds are weak.
The weak exchange is set to unity and the strong exchange to $J_r=1.90951$~\cite{ma2018anomalous}.
Following the strong-bond bipartition of Ref.~\cite{yang2026limits}, we divide the cylinder into two $L\times L$ subsystems, as illustrated in Fig.~\ref{fig:sm_qcp_lattice}.

Within subsystem $A$, the columns are indexed by $x=1,\ldots,L$ from the entanglement boundary.
Using translation symmetry along $y$, equivalent bonds are grouped into $L$ vertical and $L-1$ horizontal couplings,
\begin{equation}
    \begin{aligned}
        J_{2x-1}&:\quad (x,y)\leftrightarrow(x,y+1),
        &&x=1,\ldots,L,\\
        J_{2x}&:\quad (x,y)\leftrightarrow(x+1,y),
        &&x=1,\ldots,L-1.
    \end{aligned}
\end{equation}
Each coupling group contains $L$ bonds related by translation along $y$.
There are therefore $2L-1$ independent coupling parameters.
The first four groups are illustrated in Fig.~\ref{fig:sm_qcp_lattice}, and the same indexing convention is used in the inset of Fig.~\ref{fig:qcp_entropy}.

The learned EH uses the nearest-neighbor form of Eq.~\eqref{eq:freeform_eh}, with one variational coupling for each bond group.
Neither the LBW spatial profile nor the original exchange ratio $J_r$ is imposed during learning.
The dimerization of the original model enters through the target nearest-neighbor correlations.

\subsection{LBW reference Hamiltonian}

The LBW Hamiltonian uses the same bond structure but fixes the couplings through their distances from the entanglement cut and the exchange pattern of the original model.
In the indexing convention introduced above, the vertical-bond couplings are
\begin{equation}
    J_{2x-1}^{\rm LBW}
    =
    \frac{2\pi}{v}\left(x-\frac12\right),
    \qquad x=1,\ldots,L .
\end{equation}
For the horizontal bonds,
\begin{equation}
    J_{2x}^{\rm LBW}
    =
    \frac{2\pi}{v}\,x,
    \qquad x=1,\ldots,L-1 ,
\end{equation}
with an additional factor of $J_r$ for even $x$, corresponding to strong bonds of the original model.

We fix $v=2.697$ for all system sizes, using the value obtained for the strong-bond bipartition at the quantum critical point from the imaginary-time correlation method of Ref.~\cite{yang2026limits}.
Previous work has shown that the LBW ansatz exhibits systematic discrepancies in entanglement correlations for this bipartition, even after calibration of its overall scale~\cite{yang2026limits}.
We therefore use this calibrated LBW Hamiltonian as a reference to assess the improvement achieved by free-form learning.
The LBW $S_2$ is evaluated using the same thermal reweight-annealing procedure as for the learned EHs.

\subsection{Learning and freezing strategy}

The target correlations are obtained from the original system and held fixed throughout learning.
All variational couplings are initialized uniformly at $J_i=0.1$ and optimized at unit effective inverse temperature using the logarithmic parametrization $\theta_i=\ln J_i$.
We use the local response updates and Adam optimization described above, with a decreasing learning rate and bounded parameter updates.
No entropy data enter the optimization.

The alternating strong- and weak-bond structure produces distinct coupling scales, so the variational couplings $J_i$ can stabilize at different rates and their convergence need not follow distance from the entanglement boundary alone.
We therefore adapt the freezing sequence to the bond structure, guided by the convergence behavior of the couplings.
All couplings are initially optimized together.
As optimization proceeds, the vertical and weak horizontal coupling groups are frozen progressively from the boundary inward, followed by the strong horizontal groups.
This strategy allows different coupling groups to be optimized for different durations.
Once frozen, a coupling remains part of the simulated EH but is excluded from subsequent updates, suppressing stochastic drift.

These freezing choices facilitate optimization without prescribing a functional form for the EH couplings.
The convergence of the learned couplings, together with the final correlation matching, determines the quality of the optimization result.

\subsection{Coupling averages and entropy measurements}

For each system size, we perform ten independent learning realizations with the same target correlations.
The resulting couplings are averaged group by group to define the fixed learned EH used for entropy evaluation.
The coupling error bars in the inset of Fig.~\ref{fig:qcp_entropy} denote standard errors of the mean across these learning realizations.
We evaluate $S_{\rm vN}$ and $S_2$ for the learned EH, and $S_2$ for the LBW Hamiltonian, using the same thermal reweight-annealing procedure described above.
Their statistical uncertainties are estimated from independent Monte Carlo realizations at fixed couplings.
For the learned EH, these uncertainties do not include variations between learning realizations.
The QMC reference for $S_2$ is obtained independently from the original system using the nonequilibrium replica QMC method described above, with the same bipartition.

\end{document}